\documentclass[submitting]{nst}

\usepackage{dcolumn}
\usepackage{multirow}
\usepackage{braket}
\usepackage{float}
\usepackage[T2A,T1]{fontenc}
\usepackage[russian,english]{babel}
\usepackage{verbatim}

\usepackage{listings}
\begin{document}

\title{Calculation of microscopic nuclear level densities based on covariant density functional theory}

\author{Kun-Peng Geng}
\affiliation{%
	College of Physics, Jilin University, Changchun 130012, China
}%
\author{Peng-Xiang Du}
\affiliation{%
	College of Physics, Jilin University, Changchun 130012, China
}%
\author{Jian Li}
\email{E-mail:jianli@jlu.edu.cn}
\affiliation{%
College of Physics, Jilin University, Changchun 130012, China
}%
\author{Dong-Liang Fang}
\email{E-mail:dlfang@impcas.ac.cn}
\affiliation{%
    Institute of Modern Physics, Chinese Academy of Sciences, Lanzhou 730000, China
}%
\date{\today}

\begin{abstract}
A microscopic method for calculating nuclear level density (NLD) based on the covariant density functional theory (CDFT) is developed. The particle-hole state density is calculated by combinatorial method using the single-particle levels schemes obtained from the CDFT, and then the level densities are obtained by taking into account collective effects such as vibration and rotation. Our results are compared with those from other NLD models, including phenomenological, microstatistical and non-relativistic HFB combinatorial models. The comparison suggests that the general trends among these models are basically the same, except for some deviations from different NLD models. In addition, the NLDs of the CDFT combinatorial method with normalization are compared with experimental data, including  the observed cumulative number of levels at low excitation energy and the measured NLDs. Compared with the existing experimental data, the CDFT combinatorial method can give reasonable results.
\end{abstract}
\maketitle


\section{INTRODUCTION}

Nuclear level density (NLD) is the basic physics input for nuclear reactions. It is the key ingredient for the calculation of reaction cross sections relevant for nucleosynthesis ~\cite{{RevModPhys882},{Moller1995},{RevModPhys969},{Yal2017},{Luo2023},{Chen2023}}. The study of NLDs can date back to 1930's with Bethe’s pioneering work~\cite{PhysRev50332}. Since then, many theoretical models, such as the Back-Shifted Fermi gas model (BFM)~\cite{DILG1973269}, the Composite Constant Temperature model (CTM)~\cite{doi101139p65139} and the Generalised Superfluid model (GSM)~\cite{KONING200813}, have been adopted for the NLD study. These phenomenological models are widely used in nuclear reaction calculations. To adjust the parameters, the phenomenological models rely more or less on experimental data, however, the experimental data is limited, especially for nuclei far from the $\beta$-stability line~\cite{3}. To deal with such difficulties, the microscopic method has then been developed.

Over the past decades, various microscopic models for NLD have been proposed, from the equidistant spacing model~\cite{{WILLIAMS1971231},{19712},{OBLOZINSKY1986127},{HILAIRE1998417}}, the shell-model Monte Carlo method~\cite{{PhysRevLett834265},{PhysRevC56R1678},{PhysRevC61034303},{PhysRevC49852},{PhysRevC50836}}, the spectral distribution calculation~\cite{{PhysRevC361604},{PhysRevC51606},{PhysRevC394}}, to the independent particle model at finite temperature~\cite{{PhysRevC.96.054321},{PhysRevLett.118.022502},{PhysRevC.96.054326},{DEY2019634}}, the micro-statistical methods~\cite{{AGRAWAL1998362},{GORIELY199628},{DECOWSKI1968129},{DEMETRIOU200195}} and the random matrix method~\cite{RevModPhys.79.997}. Recently, a stochastic estimation method of level density in the configuration-interaction shell model (CISM) framework has been proposed~\cite{SHIMIZU201613} and applied to calculate the NLDs of the fission products $^{133-137}{\textrm{Xe}}$ and $^{135-138}{\textrm{Ba}}$ in Ref.~\cite{PhysRevC.107.054306}. In the past two decades, microscopic method based on the Hartree-Fock-Bogoliubov (HFB) plus combinatorial model~\cite{{3}} has been well developed. The idea of using the combinatorial method to calculate the level densities is derived from the calculation of excitation state densities~\cite{BERGER1974391}. After successfully describing the excitation state densities by means of combinations of nucleons occupying single-particle levels at the mean-field, it was a natural idea to further describe the level densities by considering the collective effects~\cite{PhysRevC.37.2600}. The combinatorial method can compete with statistical methods in reproducing experimental data, and provide the energy-, spin- and parity-dependent NLDs which are beyond the reach of statistical methods~\cite{{PhysRevC.78.064307}}. The non-relativistic Hartree-Fock-Bogoliubov combinatorial methods based on the Skyrme and Gogny effective interactions have successfully reproduced the NLDs for various nuclei ~\cite{{PhysRevC.78.064307},{PhysRevC.86.064317}} and been applied to astrophysical reactions. The accuracy of NLD is related to the basic information of nuclear structure, such as single-particle levels, deformation and binding energy. In recent years, the covariant (relativistic) density functional theory has attracted considerable attention in the nuclear physics community on account of its successful description of the complex nuclear structure and reaction dynamics~\cite{{RING1996193},{MENG2006470},{VRETENAR2005101},{NIKSIC2011519},{Meng2013},{Meng2015},{SHEN2019103713}}. For instance, it can reproduce well the isotopic shifts in the Pb isotopes~\cite{SHARMA19939} and naturally give the origin of the pseudospin and spin symmetries in the antinucleon spectrum~\cite{PhysRevLett.91.262501}, as well as provide a good description of the nuclear magnetic moments~\cite{{PhysRevC.88.064307},{JOUR}}. Recently, a microstatistical method based on CDFT has been developed to describe NLDs~\cite{PhysRevC.102.054606}. The method is applied to the calculation of NLDs of $^{94,96,98}{\textrm{Mo}}$, $^{106,108}{\textrm{Pd}}$, etc, and the NLDs are in very good agreement with experimental data over the entire energy range of measured values~\cite{PhysRevC.102.054606}. While the microstatistical method can calculate only energy-dependent NLDs, the combinatorial method can calculate the energy-, spin- and parity-dependent NLDs. Therefore, it is a meaningful attempt to calculate the NLDs using CDFT plus combinatorial method.

The theoretical framework and methods are introduced in Sec~\ref{TF}. The NLDs calculated based on CDFT combinatorial method are compared with other NLD predictions and experimental
data in Sec~\ref{RD}. Conclusions and prospects are finally drawn in Sec~\ref{SP}.
\section{THEORETICAL FRAMEWORK}\label{TF}
Covariant density functional theory starts from a Lagrangian, and the corresponding Kohn-Sham equations have the form of a Dirac equation with effective fields $S$ and $V$ derived from this Lagrangian~\cite{{osti5443763},{RING1996193},{VRETENAR2005101},{JOUR}}. Specifically, the nucleons in the nucleus are described as Dirac particles moving in the average potential field given by the meson and photon fields, interacting with each other through the exchange of meson and photon. By solving the Dirac equation:
\begin{equation}
[\boldsymbol{\alpha} \cdot{\boldsymbol{p}}+\beta(m+S)+V] \psi_{i}=\varepsilon_{i} \psi_{i},
\end{equation}
where $\varepsilon_{i}$ is the single-particle energy, and that's what we need to calculate the NLDs. The $S$ and $V$ are the relativistic scalar field, and the time-like component of vector field.

After obtaining the energy ${\varepsilon}$, spin projection ${m}$ and parity ${p}$ of the single-particle levels through CDFT, the level information is substituted into the generating function defined in the combinatorial method~\cite{3} in order to obtain the particle-hole state density $\rho_i$, and the generating function $\mathcal{Z}$ reads
\begin{equation}\label{Z1}
\mathcal{Z}(\mathit{x}_{1},\mathit{x}_{2},\mathit{x}_{3},\mathit{x}_{4})=\prod_{k=1}^{4}\prod_{i=1}^{I_{k}}(1+\mathit{x}_{k}p_{i}^{k}\mathit{y}^{\varepsilon_{i}^{k}}\mathit{t}^{m_{i}^{k}}).
\end{equation}
This generating function is a straightforward generalization of that used previously in Ref.~\cite{HILAIRE1998417} to account not only for energy ${\varepsilon_{i}^{k}}$ but also for spin projection ${m_{i}^{k}}$ and parity $p_{i}^{k}$. The variables ${x}_{k}$, ($k$ = 1, ..., 4) enable us to count the number of particles and holes, $y$ enables us to keep track of the excitation energies and $t$ of the spin projections, the indexes ${I}_{1}$,...,${I}_{4}$ denote the number of discrete states considered for each set of single-particle and single-hole states. In practice, ${I}_{2}$ = $Z$ and ${I}_{4}$ = $N$ for a nucleus with $Z$ protons and $N$ neutrons. $\mathcal{Z}$ can be expanded in powers of $\mathit{x}_{k}$ writing
\begin{equation}\label{Z2}
  \mathcal{Z}(\mathit{x}_{1},\mathit{x}_{2},\mathit{x}_{3},\mathit{x}_{4})=\sum_{\mathcal{N}}\mathcal{F}_{\mathcal{N}}(\mathit{y},\mathit{t})\prod_{k=1}^{4}\mathit{x}_{k}^{N_{k}},
\end{equation}
the symbol $\mathcal{N}$ denoting again any integers combination $(N_{1},N_{2},N_{3},N_{4})$. The function $\mathcal{F}({\mathit{y},\mathit{t}})$ can be expanded into powers of $y$ and $t$ writing
\begin{equation}\label{Z3}
  \mathcal{F}_{\mathcal{N}}(\mathit{y},\mathit{t})=\sum_{U}\sum_{M}\sum_{P=-1,+1}\mathit{C}_{\mathcal{N}}(U,M,P)\mathit{y}^{U}\mathit{t}^{M},
\end{equation}
where the $U$, $M$ and $P$ are the excitation energy, spin projection and parity. The coefficients $\mathit{C}_{\mathcal{N}}(U,M,P)$ are the numbers of solutions. Through the coefficients $\mathit{C}_{\mathcal{N}}(U,M,P)$, the simplest definition for particle-hole state density $\mathit{\rho}_{i}$ reads
\begin{equation}\label{Z3}
  \mathcal{\rho}_{i}(\mathit{U},\mathit{M},\mathit{P})=\frac{1}{\varepsilon_0}\mathit{C}_{\mathcal{N}}(U,M,P),
\end{equation}
but the state density $\mathit{\rho}_{i}$ turn out to be strongly dependent on any unit energy $\varepsilon_0$. Therefore, another method suggested by Williams~\cite{WILLIAMS1972225} is employed in this paper to limit the discretization effects as explained below.

Summing all the $\mathit{C}_{\mathcal{N}}$ up to a given excitation energy $U$, we first obtain the cumulated number of states $N_{\mathcal{N}}(U, M, P)$ which represents the number of particle-hole states with excitation energy $E$ such that $0\leq{E}<U$. The particle-hole state density defined as
\begin{equation}\label{Z3}
  \mathcal{\rho}_{i}=\frac{dN_{\mathcal{N}}(U,M,P)}{dU}.
\end{equation}

The particle-hole state density $\rho_i$ calculated is only related to the particle-hole excitation. Two special collective effects need to be taken into account in order to obtain the level density, including rotation and vibrational enhancement. If the nucleus under consideration displays spherical symmetry, the intrinsic and laboratory frames coincide, and the level density is trivially obtained through the relation~\cite{PhysRev.50.332}
\begin{equation}\label{{sph}}
  \rho_{\rm{sph}}(U,M,P) = \rho_{i}(U,M=J,P) - \rho_{i}(U,M=J+1,P).
\end{equation}

For deformed nuclei, within the axial symmetry hypothesis, the NLD after accounting for rotational effect reads
\begin{equation}\begin{split}
   \rho_{\rm{def}}(U,M,P)& ={\dfrac{1}{2}[ \sum_{K=-J,K\neq0}^{J}\rho_{i}(U-E^{J,K}_{\rm{rot}} ,K,P)]}\\
    &\quad +(\delta_{(J \rm{even})}\delta_{(p=+)}\rho_{i}(U-E^{J,K}_{\rm{rot}} ,0,P)\\
    &\quad + \delta_{(J \rm{odd})}\delta_{(p = -)}\rho_{i}(U-E^{J,K}_{\rm{rot}} ,0,P)).
\end{split}\end{equation}
Factor 1/2 accounts for the fact that in mirror axially symmetric nuclei, the intrinsic states with spin projections $+K$ or $-K$ give rise to the same rotational levels. Moreover, in the second term of the summation, the symbol $\delta_{(x)}$ (defined by $\delta_{(x)}=1$ if $x$ holds true, and 0 otherwise) restricts the rotational bands built on intrinsic states with spin projection $K=0$ and parity $P$ to the levels sequences 0, 2, 4,... for $P=+$ and 1, 3, 5,... for $P=-$. Finally, the rotational energy is obtained with the well-known expression~\cite{DOSSING1974493}
\begin{equation}\label{rot}
  	E^{J,K}_{\rm{rot}} = \dfrac{J(J+1)-K^{2}}{2\mathcal{J}_{\bot}},
\end{equation}
where $ \mathcal{J}_{\bot} $ is the moment of inertia of a nucleus rotating around an axis perpendicular to the symmetry axis. In this article $ \mathcal{J}_{\bot} $ is approximated by the rigid-body value $\mathcal{J}_\perp^{\rm{rigid}}$ which reads
\begin{equation}\label{rigid}
  \mathcal{J}_\perp^{\rm{rigid}} = \dfrac{2}{5}mR^{2}(1+\sqrt{\dfrac{5}{16\pi}}\beta^{2}),
\end{equation}
for an ellipsoidal shape with quadrupole deformation parameter $\beta$. The vibration enhancement is approximated by equation~\cite{HILAIRE200663}
\begin{equation}\label{vib}
  	K_{\rm{vib}} = \rm{exp}[\delta S - (\delta U/T)],
\end{equation}
where $S$ is the entropy, $U$ is the excitation energy, and $T$ is the nuclear temperature.

Finally, a phenomenological damping function is introduced to avoid sharp transitions between spherical and deformed level densities affecting the NLD predictions. The expression of the NLD after introducing the damping function can be written as~\cite{DEMETRIOU200195}
\begin{equation}\begin{split}
	\rho(U,M,P)& = [1 - f_{\rm{dam}}(U)]K_{\rm{vib}}\rho_{\rm{sph}}(U,M,P)\\
     &\quad+ f_{\rm{dam}}(U)K_{\rm{vib}}\rho_{\rm{def}}(U,M,P).
\end{split}\end{equation}
The damping function $f_{\rm{dam}}$ is expressed as~\cite{PhysRevC.86.064317}
\begin{equation}
  f_{\rm{dam}}=1-\frac{1}{1+\exp{({\beta}-0.18)/0.038}},
\end{equation}
where $\beta$ is the quadrupole deformation parameter. Parameters 0.18 and 0.038 have been adjusted according to the experimental data at the neutron separation energy $S_n$~\cite{1}. The expression of $f_{\rm{dam}}$ has been evolved in Ref.~\cite{PhysRevC.86.064317}, which has been simplified to depend only on the nuclear deformation thus reducing the number of phenomenological parameters. This damping function is used to suppress discontinuities that occur between spherical and deformed NLDs and ensure a smooth shape change from deformed to spherical.

\section{RESULTS AND DISCUSSION}\label{RD}

In this section, we present our results of the NLDs from the combinatorial method based on CDFT and compare them with the results from other NLD models~\cite{KONING20122841} and experiments~\cite{PhysRevC.87.014319, PhysRevC.68.064306}. The effective meson-exchange interaction parameter PK1~\cite{1999} is adopted throughout the CDFT calculations. For spherical CDFT calculations, we fix the box size $R_{\rm{box}}$ = 20 fm, and step size ${\Delta}r$ = 0.1 fm. In the present deformed CDFT calculations, both the Dirac equation for
nucleons and the Klein–Gordon equations for mesons are
solved in an isotropic harmonic oscillator basis and a
basis of 18 major oscillator shells is adopted.
\begin{figure}[h]
  \centering
  \includegraphics[width=8cm]{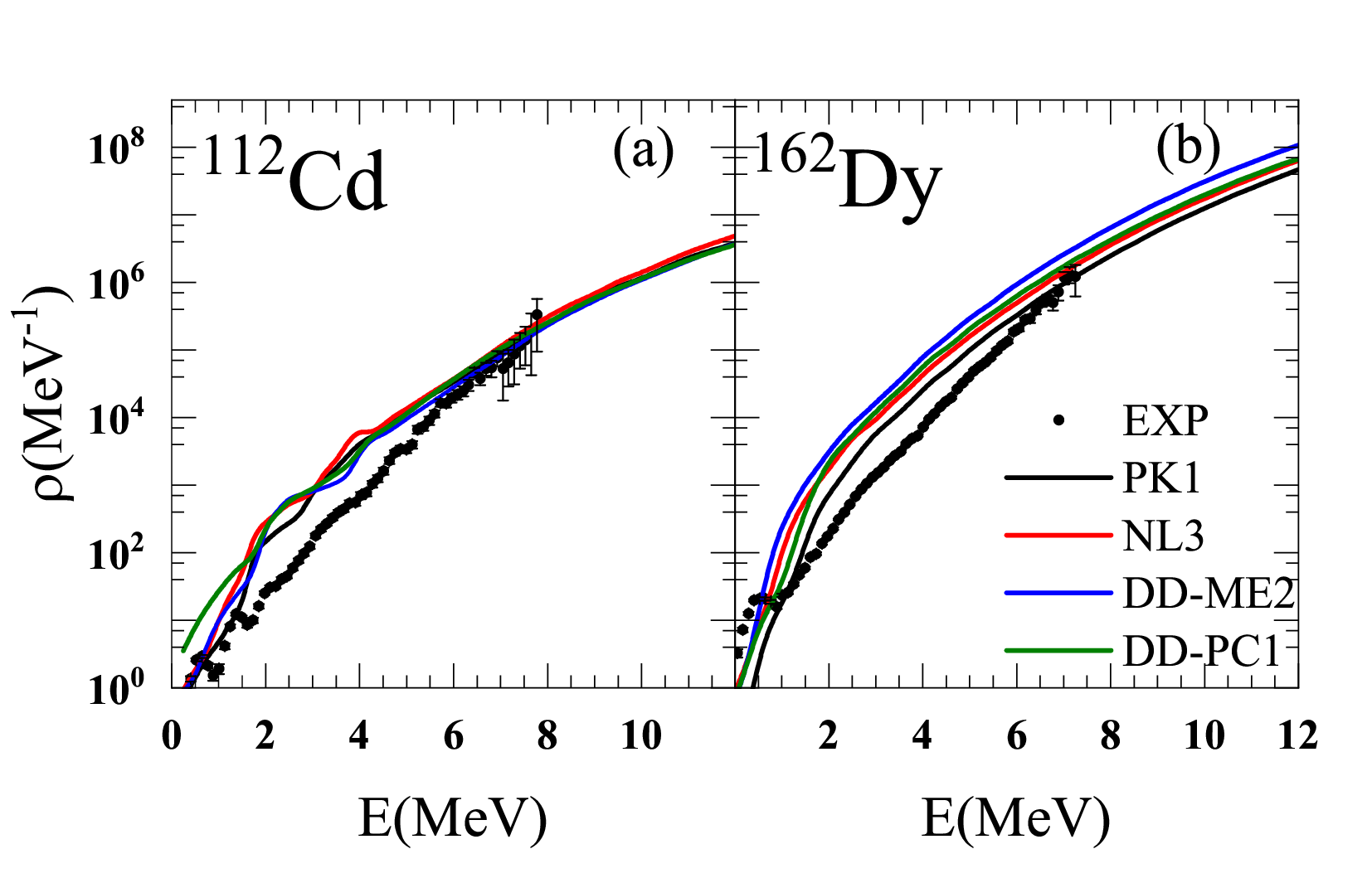}
  \caption{(Color online) Comparison of NLDs calculated using different effective interactions under CDFT with experimental data~\cite{{PhysRevC.87.014319}, {PhysRevC.68.064306}}.}.\label{cd112ef}
\end{figure}

As the combinatorial results rely on the single-particle levels properties. To understand the effect of using different effective interactions on the NLDs, the NLDs of $^{112}$Cd and $^{162}$Dy obtained with different CDFT effective interactions PK1~\cite{1999}, NL3~\cite{LALAZISSIS200936}, DD-ME2~\cite{NIKSIC20141808} and DD-PC1~\cite{PhysRevC.71.024312} are presented for the sake of comparison in Fig.~\ref{cd112ef}. For nuclei with small deformation, $^{112}$Cd, the NLDs for four effective interactions are close for excitation energy above 4 MeV. However, a significant difference is observed for excitation energy below 4 MeV. The difference comes from the fact that the single-particle energies near the vicinity of the Fermi level are very sensitive to the choice of effective interactions especially when the Fermi level is near the proton or neutron shells. Meanwhile, CDFT calculations with chosen interactions do not reproduce the single-particle levels of the magic nuclei $^{132}$Sn~\cite{PhysRevC.84.014305} very well, which explains the deviations of our results from the measurements in Fig.\ref{cd112ef}. For the well deformed $^{162}{\textrm{Dy}}$, the NLDs calculated with four effective interactions deviate from each other for the whole region of excitation energy, although the overall trend is consistent. This deviation at low excitation energy is mainly caused by differences in the single-particle energies around the Fermi level. Meanwhile, the entropy $S$ obtained from the four effective interactions is significantly different for the nuclei with large deformation, and ultimately leading to the deviation of NLDs at high excitation energy. Compared with the experimental data of $^{112} {\textrm{Cd}}$~\cite{PhysRevC.87.014319} and $^{162}{\textrm{Dy}}$~\cite{PhysRevC.68.064306} extracted by the Oslo group from the analysis of particles-$\gamma$ coincidence in the ($^{3}{\textrm{He}}$, $\alpha\gamma$) and ($^{3}{\textrm{He}}$, $^{3}{\textrm{He}}$'$\gamma$) reactions, the deviation between the NLDs obtained by using PK1 effective interaction and the results from other effective interactions is within a reasonable range.

\begin{figure}[h]
  \centering
  \includegraphics[width=8cm]{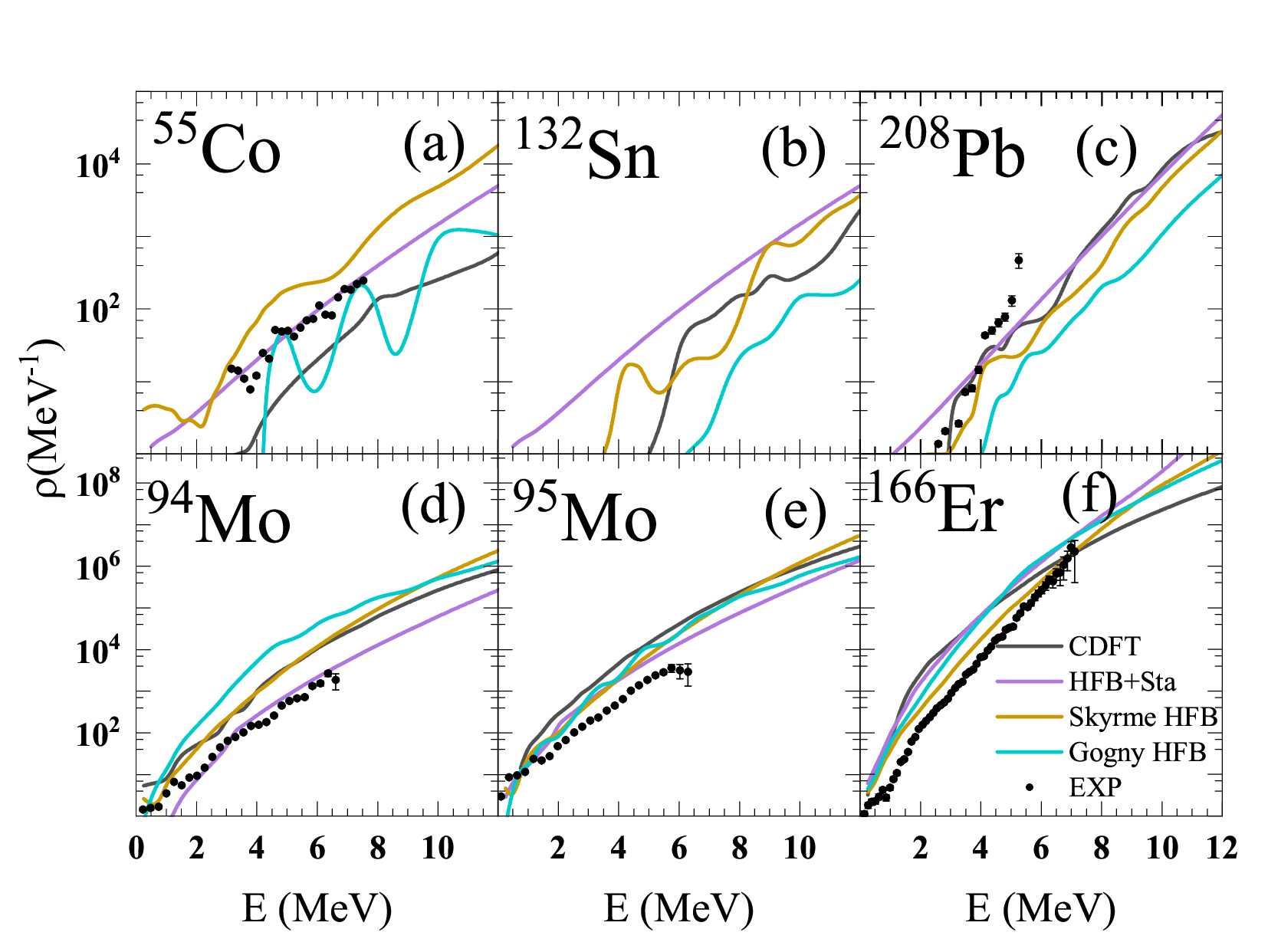}
  \caption{(Color online) Comparison of NLDs calculated based on CDFT combinatorial method with results from HFB combinatorial methods and microstatistical method (HFB+Sta)~\cite{{PhysRevC.78.064307}, {PhysRevC.86.064317},{DEMETRIOU200195}}. The experimental data are taken from Refs.~\cite{{PhysRevC.88.064324},{PhysRevC.73.034311},{PhysRevC.79.024316},{PhysRevC.63.044309}}. }\label{co55hf}
\end{figure}

As the non-relativistic HFB combinatorial methods, including Skyrme and Gogny interactions, have been widely used in NLD predictions~\cite{HILAIRE200663, PhysRevC.87.014319}, the NLDs of spherical nuclei ( $^{55}{\textrm{Co}}$, $^{132}{\textrm{Sn}}$, $^{208}{\textrm{Pb}}$ ) and deformed nuclei ( $^{94}{\textrm{Mo}}$, $^{95}{\textrm{Mo}}$, $^{166}{\textrm{Er}}$ ) calculated from the present CDFT combinatorial method are given in Fig.~\ref{co55hf} and compared with them and the corresponding experimental data.

In $^{55}{\textrm{Co}}$, the CDFT combinatorial method gives a lower NLD compared to the experimental data, while the Skyrme combinatorial method gives a higher NLD. Although the result of Gogny combinatorial method is closer to the experimental data to a certain extent, there is a strong oscillation. For $^{132}{\textrm{Sn}}$ and $^{208}{\textrm{Pb}}$, the results of CDFT combinatorial method are similar to that of Skyrme combinatorial method, but the results of Gogny combinatorial method are significantly lower than the two methods. Compared with the experimental data, the CDFT combinatorial method can better describe the NLD of $^{208}{\textrm{Pb}}$. The NLDs of $^{55}{\textrm{Co}}$, $^{132}{\textrm{Sn}}$ and $^{208}{\textrm{Pb}}$ from the three combinatorial methods are more or less oscillatory, which is because the three nuclei are spherical nuclei with highly degenerate single-particle levels, and there is a strong shell effect.

For $^{94}{\textrm{Mo}}$, both the results of the CDFT and Skyrme combinatorial methods are closer to the experimental data than the Gogny combinatorial method. For $^{95}{\textrm{Mo}}$ and $^{166}{\textrm{Er}}$, the NLDs obtained by the three combinatorial methods are similar and slightly larger than the experimental data. It should be emphasized that the deformed nuclei $^{94}{\textrm{Mo}}$, $^{95}{\textrm{Mo}}$ and $^{166}{\textrm{Er}}$ break single-particle level degeneracy, thus obtain a smoother NLDs as shown in Fig.~\ref{co55hf}. It is worth noting that the CDFT combinatorial method gives higher NLDs for $^{94}{\textrm{Mo}}$ and $^{95}{\textrm{Mo}}$. One possible reason is that the CDFT combinatorial method uses rigid-body value when considering rotational effect, which is not appropriate for soft nuclei~\cite{PhysRevC.101.054315}.

In addition, the NLDs of $^{55}{\textrm{Co}}$ and $^{94}{\textrm{Mo}}$ are well described by the microstatistical model~\cite{DEMETRIOU200195} relative to the CDFT combinatorial method, while the description of the NLDs of $^{95}{\textrm{Mo}}$ and $^{166}{\textrm{Er}}$ are similar between these two models. In comparison, microstatistical model gives relatively smooth results for spherical and deformed nuclei, which do not well reflect shell effect. Therefore, it can be obtained that the CDFT combinatorial method has a similar ability to describe NLDs as the other two combinatorial methods and microstatistical model.

\begin{figure}[h]
  \centering
  \includegraphics[width=8cm]{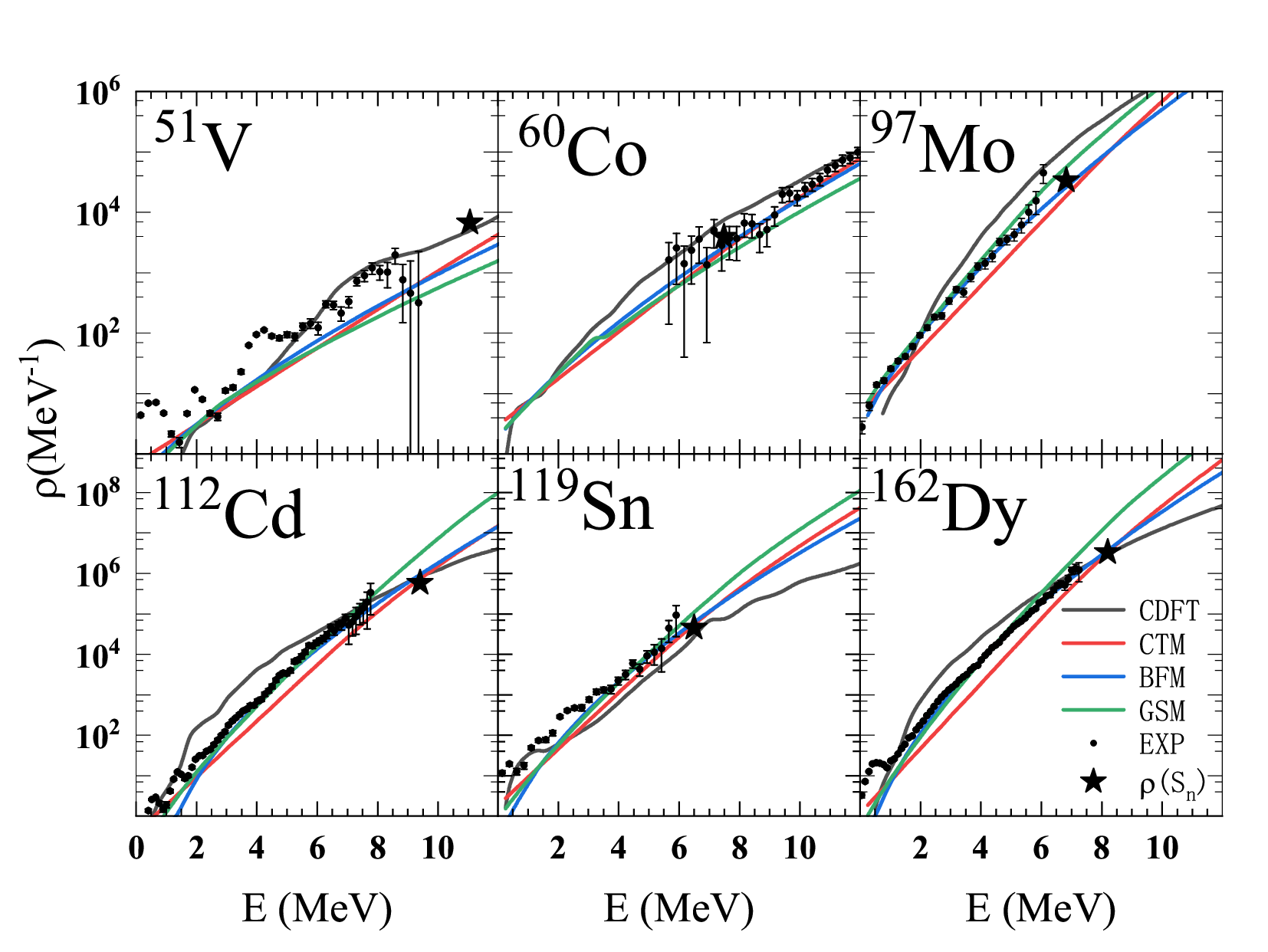}
  \caption{(Color online) Comparison of NLDs obtained from CDFT combinatorial method with other NLD predictions~\cite{KONING20122841} and experimental data~\cite{{PhysRevC.73.064301},{PhysRevC.73.034311},{PhysRevC.87.014319},{PhysRevC.87.014319},{PhysRevC.68.064306},{PhysRevC.68.064306}}. The full asterisk corresponds to the experimental data at the neutron separation energy $S_n$~\cite{1}.}\label{v51s}
\end{figure}

At the same time, it's also necessary to compare with phenomenological models (BFM, CTM and GSM)~\cite{KONING20122841} that have been widely applied. In particular, the phenomenological models can well describe the experimental data at the neutron separation by fitting experimental data. The level densities at the neutron separation energy are the most commonly used experimental data for research and are obtained by the mean distance $D_0$ of the $s$-wave resonance. The NLDs of even-even nuclei ($^{112}{\textrm{Cd}}$, $^{162}{\textrm{Dy}}$), odd-$A$ nuclei ($^{51}{\textrm{V}}$, $^{97}{\textrm{Mo}}$, $^{119}{\textrm{Sn}}$) and odd-odd nucleus ($^{60}{\textrm{Co}}$) calculated from the present CDFT combinatorial method are given in Fig.~\ref{v51s} and compared with results of phenomenological models and the available experimental data. It can be seen from Fig.~\ref{v51s} that the CDFT combinatorial method has similar ability to the phenomenological models in reproducing the experimental data on neutron separation energy, especially the CDFT combinatorial method can describe the NLD of $^{51}{\textrm{V}}$ well. In addition, the overall description of experimental data below neutron separation energy by the CDFT combinatorial method is also similar to that of the phenomenological methods. In conclusion, while there are some differences in the results for each NLD model, the overall ability to describe experimental data is similar.
\begin{figure}[h]
  \centering
  \includegraphics[width=9cm]{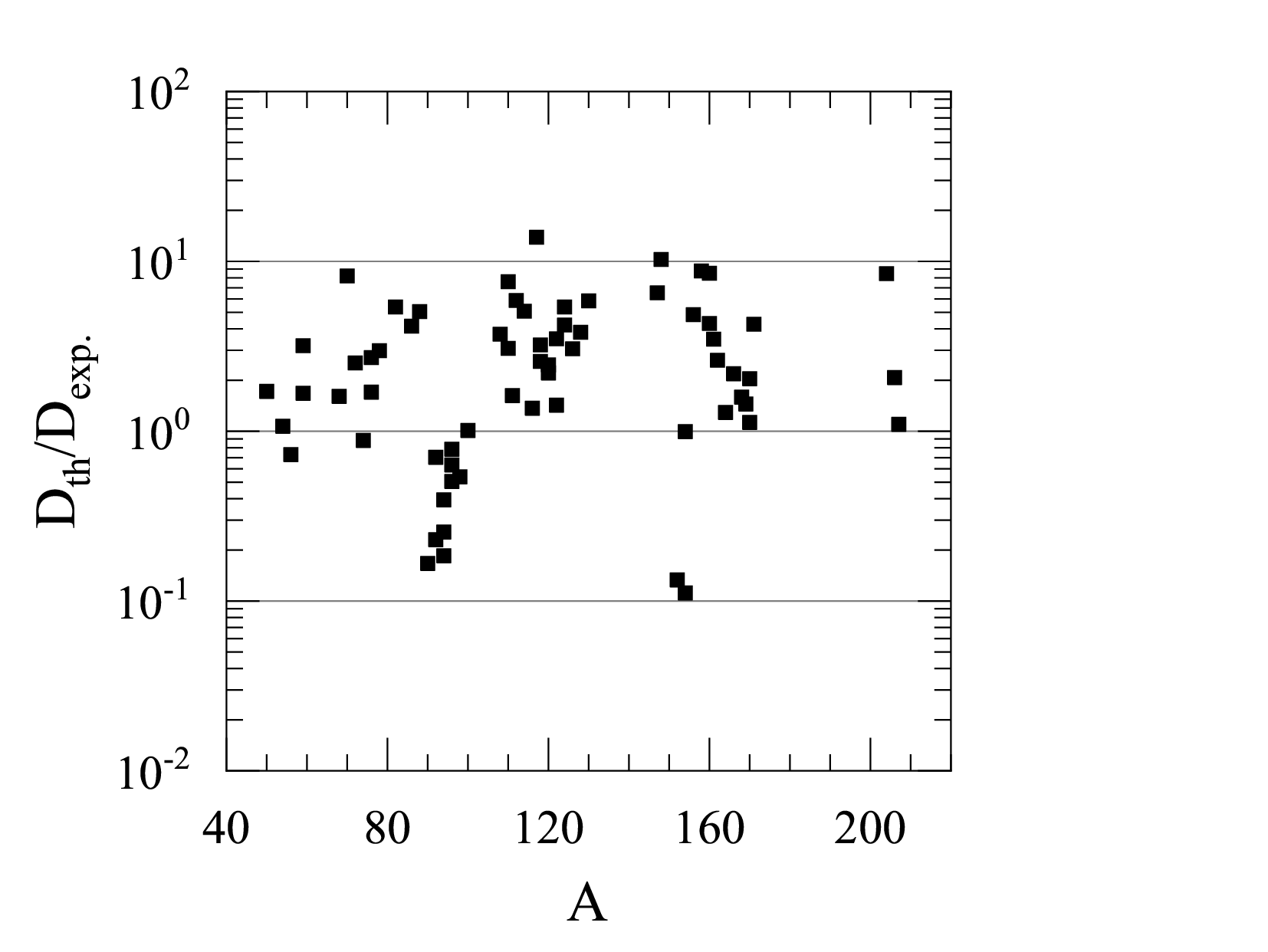}
  \caption{The ration of CDFT combinatorial method ($D_{\rm{th}}$) to the experimental ($D_{\rm{exp.}}$) s-wave neutron resonance spacings compiled in Ref.~\cite{1}.}\label{frms}
\end{figure}

The most extensive and reliable source of experimental information on NLD remains the s-wave neutron resonance spacings $D_0$~\cite{1} and the observed low-energy excited levels~\cite{1}. To measure the dispersion between theoretical and experimental $D_0$, the $f_{\rm{rms}}$ factor is defined as
\begin{equation}
f_{\rm{rms}}={\rm{exp}}{\left[\frac{1}{N_e}\sum^{N_e}_{i=1}{\rm{ln}^2\frac{D^{i}_{\rm{th}}}{D^{i}_{exp.}}}\right]}^{1/2},
\end{equation}
where $D_{\rm{th}}$($D_{\rm{exp.}}$) is the theoretical (experimental) resonance spacing and $N_e$ is the number of nuclei in the compilation. The results for the ratios of $D_{\rm{th}}$ and $D_{\rm{exp.}}$ are shown in Fig.~\ref{frms}, and the $f_{\rm{rms}}=3.62$, a total of 66 nuclei are counted. Under the same calculation conditions, the $f_{\rm{rms}}$ of Gogny (D1S) interaction based on HFB plus combinatorial method is 7.25~\cite{3}. At present, the result of CDFT combinatorial method is bigger than $f_{\rm{rms}}=1.8$ deviation of the phenomenological back-shifted Fermi gas model~\cite{KONING200813}and the $f_{\rm{rms}}=2.1$ value obtained with microstatistical method~\cite{DEMETRIOU200195}. But the HFB combinatorial based on Skyrme effective interaction model improved pairing correlation~\cite{HILAIRE200663} and collective effects, then get the $f_{\rm{rms}}=2.3$~\cite{{PhysRevC.78.064307}}. In subsequent work, it is expected to improve the CDFT combinatorial method in the same way.
\begin{figure}[h]
  \centering
  \includegraphics[width=8cm]{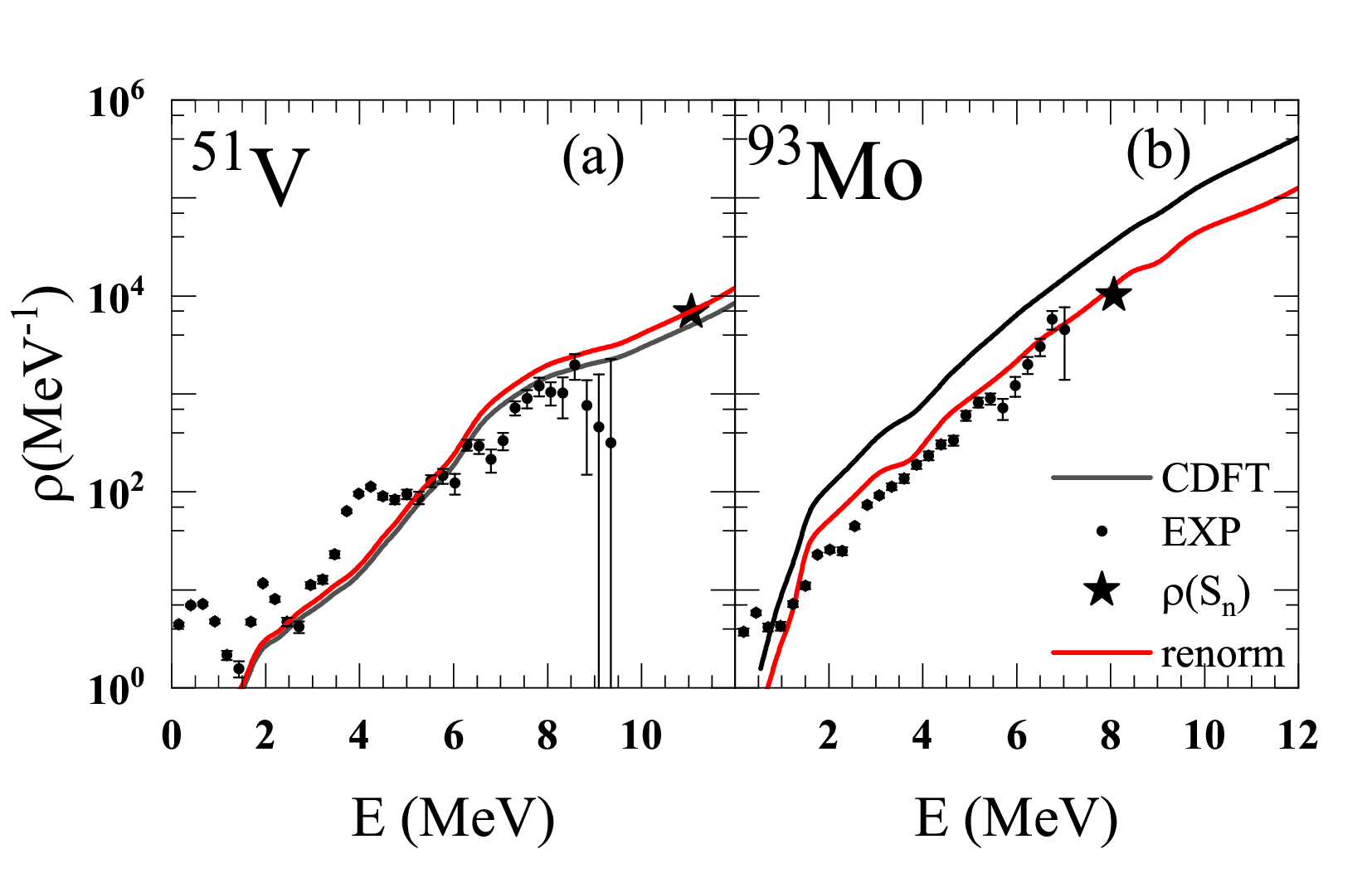}
  \caption{(Color online) The NLDs obtained from CDFT combinatorial method with (red) and without (black) normalization. The experimental data are taken from Refs.~\cite{{PhysRevC.73.064301},{PhysRevC.73.034311},{1}}.}\label{v51re}
\end{figure}

When phenomenological NLDs are used in nuclear physics applications, such as nuclear data evaluation or accurate and reliable estimation of reaction cross sections, one often plays with the few parameters on which phenomenological expressions depend~\cite{Alhassan2022}. The results of the combinatorial method also can normalize both the experimental level scheme at low excitation energy and the neutron resonance spacings at $U=S_n$ in a way similar to what is usually done with analytical formulas. More precisely, the normalized level density can be obtained by
expression~\cite{PhysRevC.78.064307}

\begin{equation}
\rho(U,J,P)_{}=e^{\alpha\sqrt{U-\delta}}\times{\rho(U-\delta,J,P)},
\end{equation}
where the energy shift $\delta$ is essentially extracted from the analysis of the cumulative number of levels, and the $\alpha$ can be obtained through the expression
\begin{equation}
\rho_{\rm{th}}(S_n)\times e^{\alpha\sqrt{S_n}}=\rho_{\rm{exp.}}(S_n).
\end{equation}
With such a normalization, the experimental low-lying states and the $D_0$ values can be reproduced reasonably well.

\begin{figure}[h]
  \centering
  \includegraphics[width=8cm]{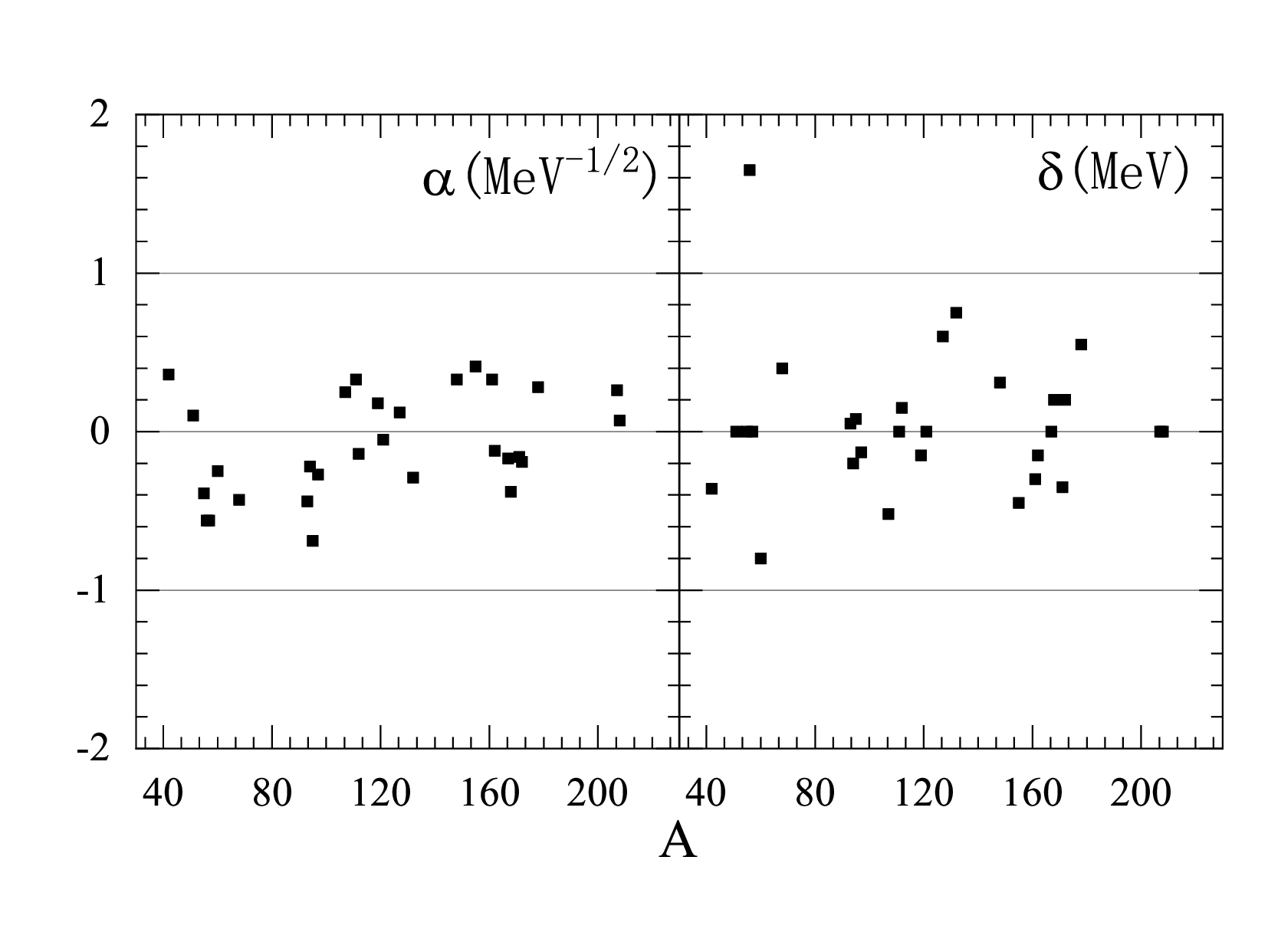}
  \caption{Normalized parameters $\alpha$ (left) and $\delta$ (right) values are plotted as a function of the atomic mass number.}\label{alpha}
\end{figure}
As an illustration, the variation of NLDs obtained by the CDFT combinatorial method before and after normalization is shown in Fig.~\ref{v51re}. When the normalization is applied, the NLDs will pass through the experimental data at the neutron separation energy $S_n$. As shown in Fig.~\ref{v51re}(a), if the theoretical calculation value is close to the experimental data at the neutron separation energy $S_n$, the results before and after normalization will not change much. In Fig.~\ref{v51re}(b), the results after normalization will be in better agreement with the experimental data ~\cite{{PhysRevC.73.064301}}, especially above 2 MeV. This is because the coefficient $e^{\alpha\sqrt{U-\delta}}$ is related to the excitation energy $U$, and when $U$ is small, the normalization has little effect. Some nuclear $\alpha$ and $\delta$ values are shown in Fig.~\ref{alpha}.

\begin{figure*}
  \centering
  \includegraphics[width=16cm]{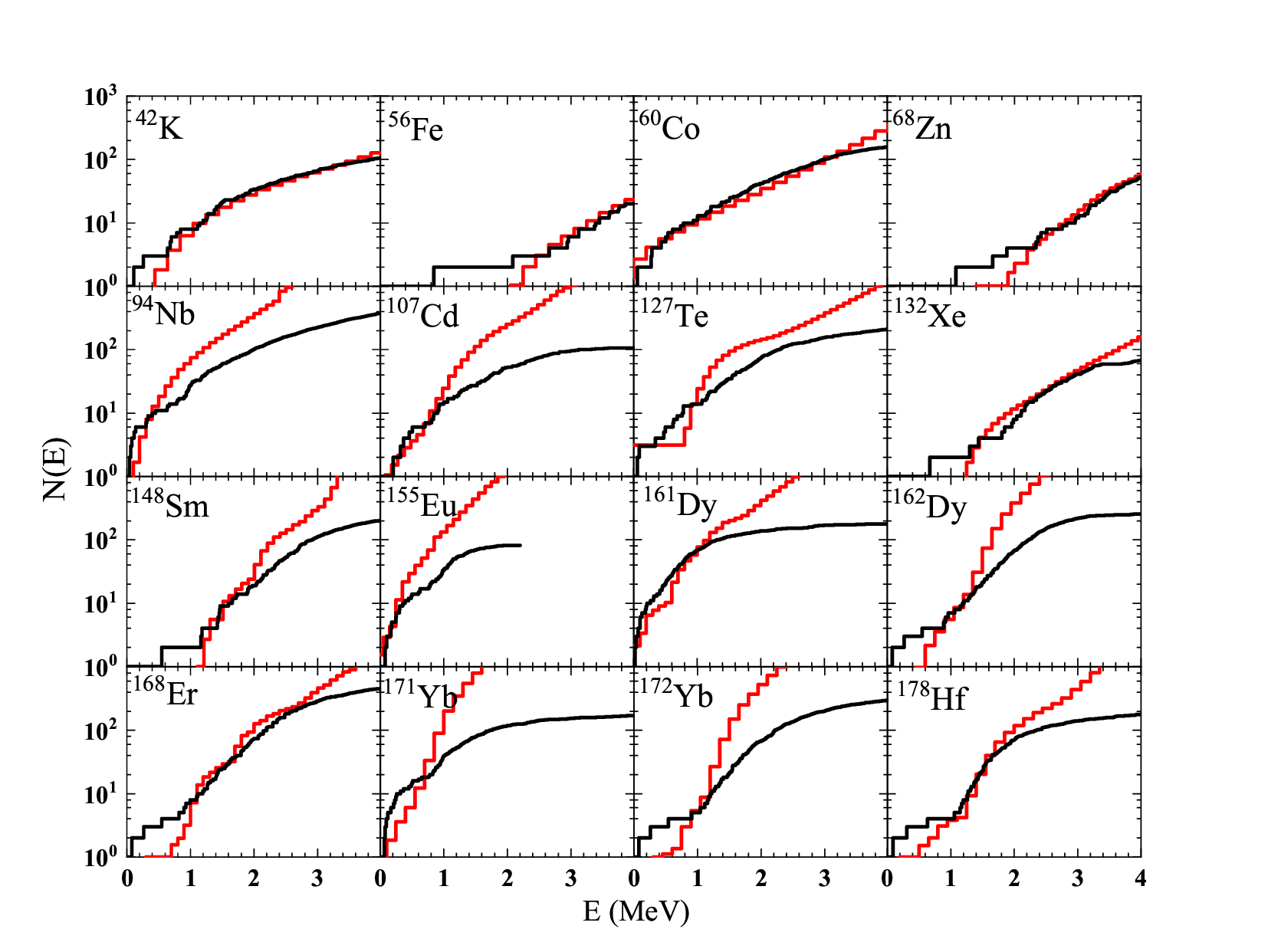}
  \caption{(Color online) Comparison of the results obtained from the CDFT combinatorial method (red lines) with the cumulative number of observed levels~\cite{1} (black lines) as a function of the excitation energy.}\label{k42}
\end{figure*}

Finally, the NLDs calculated based on the CDFT combinatorial method are compared with the observed low-energy excited levels, which are the most extensive and reliable source of experimental information on NLDs~\cite{1}. The cumulative number of nuclear levels $N(U)$ indicates the sum of the number of all levels below some excitation energy $U$ (including $U$)
\begin{equation}
N(U)={\sum_M}{\sum_P}N(U,M,P),
\end{equation}
where $N(U,M,P)$ is the cumulative number of levels under spin $M$ and parity $P$ to the excitation energy $E$. The predicted cumulative number of levels $N(U)$ are compared with the experimental data~\cite{1} in Fig.~\ref{k42}, including light and heavy as well as spherical and deformed nuclei. Globally, the results of the CDFT combinatorial method are in reasonably good agreement with experimental data at lower excitation energy, especially for light nuclei. At high excitation energy, the theoretically calculated cumulative number of nuclear levels is higher than the experimental value. The NLDs increase rapidly with the increase of excitation energy, and the $N(U)$ should also gradually increase at high excitation energy. However, the experimental levels of the cumulative number gradually stabilized, which may be limited by the experimental conditions and cannot completely count the number of levels.

\begin{figure*}
  \centering
  \includegraphics[width=16cm]{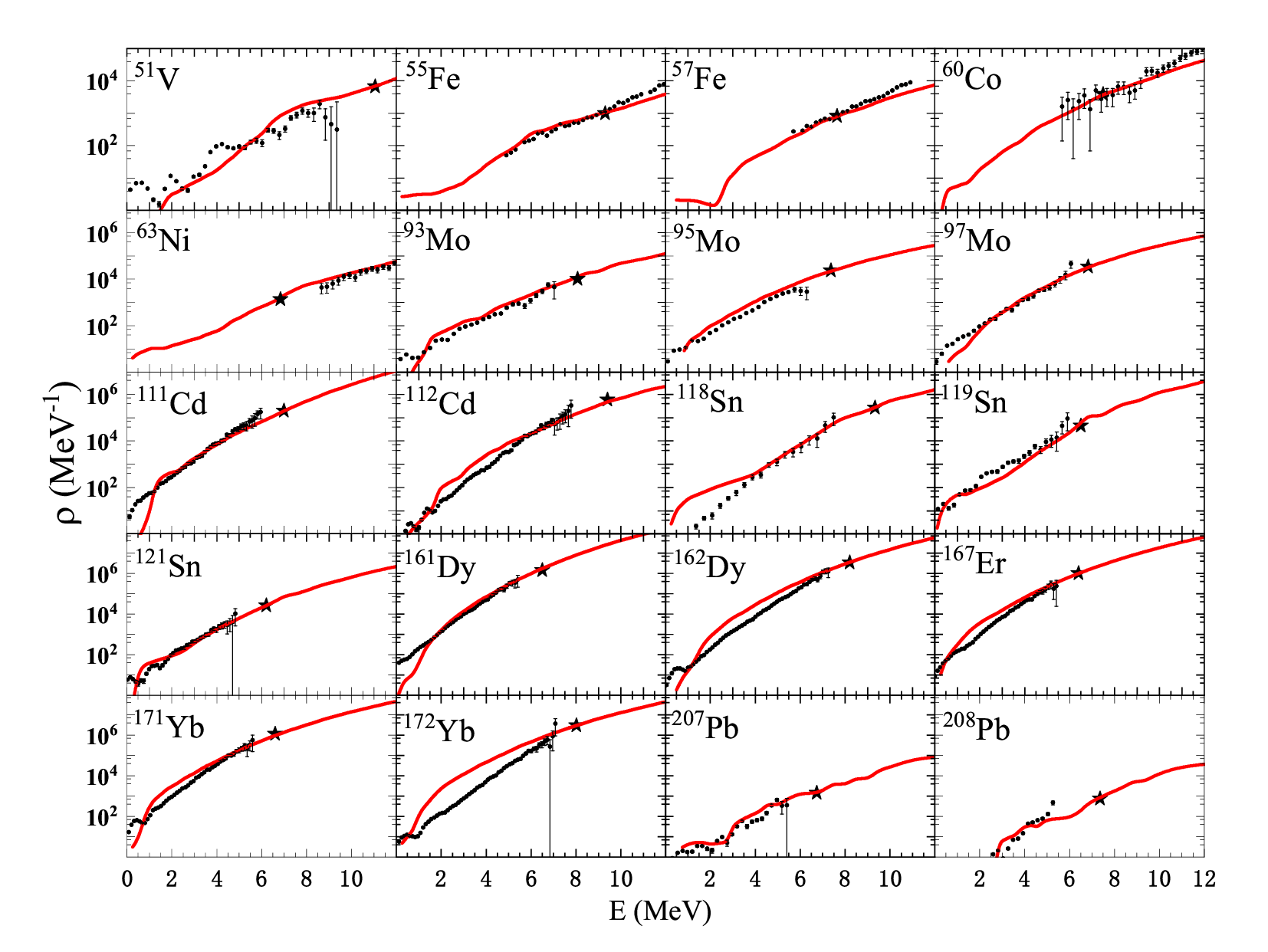}
  \caption{(Color online) Comparison between the calculated NLDs based on the CDFT combinatorial method (red lines) and the experimental data~\cite{{PhysRevC.73.064301},{PhysRevC.92.014303},{PhysRevC.80.034305},{PhysRevC.73.034311},{PhysRevC.87.014319},{PhysRevC.81.064311},{PhysRevC.68.064306},{PhysRevC.63.044309},{PhysRevC.70.054611},{PhysRevC.79.024316}} (black dot). The full asterisk corresponds to the experimental data at the neutron separation energy $S_n$~\cite{1}. }\label{v51ld}
\end{figure*}

In Fig.~\ref{v51ld}, the CDFT combinatorial method predictions after normalization are compared with the experimental data extracted by the Oslo group ~\cite{{PhysRevC.73.064301},{PhysRevC.73.034311},{PhysRevC.87.014319},{PhysRevC.81.064311},{PhysRevC.68.064306},{PhysRevC.63.044309},{PhysRevC.70.054611},{PhysRevC.79.024316}} and the particle evaporation spectrum~\cite{{PhysRevC.92.014303},{PhysRevC.80.034305}}. The Oslo method is model-dependent. In order to extract the absolute value of the total level density from the measured data, the so-called experimental NLDs need to be normalized by the total level density at the neutron binding energy, which in turn is derived from the neutron resonance spacing. For a meaningful comparison between the CDFT combinatorial predictions and the Oslo group data, it is therefore important to normalize the NLDs of the CDFT combinatorial method to the level density value at $U=S_n$ considered by the Oslo group. As shown in Fig.~\ref{v51ld}, the results of the CDFT combinatorial method after normalization agree well with the experimental data below $S_n$, except for the small NLDs of ${^{111}\textrm{Cd}}$ and ${^{161}\textrm{Dy}}$ at low excitation energy. The smaller result is due to the larger energy spacing of the theoretically calculated single-particle levels near the Fermi level. Overall, the results obtained by the CDFT combinatorial method are reliable.

\section{SUMMARY AND PROSPECTS}\label{SP}
The combinatorial method has been adopted to describe the nuclear level densities for nuclear reaction calculations. The particle-hole state density is obtained with a combinatorial method using the single-particle level provided by the CDFT based PK1 effective interaction. After accounting of collective effects, including vibration and rotation, the energy-, spin-, and parity-dependent NLDs are obtained. Our results are compared with those from other NLD models, including phenomenological, microstatistical as well as non-relativistic HFB combinatorial models. The comparison suggests that besides some small deviation from different NLD models, the general trends among these models are basically the same. In conclusion, the CDFT combination method is capable to reproduce the experimental data at or below the neutron separation energy. It implies that the CDFT combinatorial method is as reliable as other models to describe NLDs. Finally, the NLDs of the CDFT combinatorial method with normalization are compared with experimental data, obtaining excellent agreement with the observed cumulative number of levels at low excitation energy and the measured NLDs below the neutron separation energy.

Our results exhibit the predictive power of the CDFT combinatorial method, even though the pairing correlations are not considered while the collective effects are empirical. In our following work, the CDFT combinatorial method can be improved by considering the inclusion of energy-dependent pairing correlations and taking a partition function approach to the treatment of vibrational enhancement. These results will surely help our study on important neutron capture process such as r-process.

\begin{acknowledgments}
This work was supported by the Natural Science Foundation of Jilin Province (No. 20220101017JC) and the National Natural Science Foundation of China (No. 11675063) as well as the Key Laboratory of Nuclear Data Foundation (JCKY2020201C157).
\end{acknowledgments}

\clearpage
\bibliography{Refs}

\begin{thebibliography}{75}%
\makeatletter
\providecommand \@ifxundefined [1]{%
 \@ifx{#1\undefined}
}%
\providecommand \@ifnum [1]{%
 \ifnum #1\expandafter \@firstoftwo
 \else \expandafter \@secondoftwo
 \fi
}%
\providecommand \@ifx [1]{%
 \ifx #1\expandafter \@firstoftwo
 \else \expandafter \@secondoftwo
 \fi
}%
\providecommand \natexlab [1]{#1}%
\providecommand \enquote  [1]{``#1''}%
\providecommand \bibnamefont  [1]{#1}%
\providecommand \bibfnamefont [1]{#1}%
\providecommand \citenamefont [1]{#1}%
\providecommand \href@noop [0]{\@secondoftwo}%
\providecommand \href [0]{\begingroup \@sanitize@url \@href}%
\providecommand \@href[1]{\@@startlink{#1}\@@href}%
\providecommand \@@href[1]{\endgroup#1\@@endlink}%
\providecommand \@sanitize@url [0]{\catcode `\\12\catcode `\$12\catcode
  `\&12\catcode `\#12\catcode `\^12\catcode `\_12\catcode `\%12\relax}%
\providecommand \@@startlink[1]{}%
\providecommand \@@endlink[0]{}%
\providecommand \url  [0]{\begingroup\@sanitize@url \@url }%
\providecommand \@url [1]{\endgroup\@href {#1}{\urlprefix }}%
\providecommand \urlprefix  [0]{URL }%
\providecommand \Eprint [0]{\href }%
\providecommand \doibase [0]{http://dx.doi.org/}%
\providecommand \selectlanguage [0]{\@gobble}%
\providecommand \bibinfo  [0]{\@secondoftwo}%
\providecommand \bibfield  [0]{\@secondoftwo}%
\providecommand \translation [1]{[#1]}%
\providecommand \BibitemOpen [0]{}%
\providecommand \bibitemStop [0]{}%
\providecommand \bibitemNoStop [0]{.\EOS\space}%
\providecommand \EOS [0]{\spacefactor3000\relax}%
\providecommand \BibitemShut  [1]{\csname bibitem#1\endcsname}%
\let\auto@bib@innerbib\@empty
\bibitem [{\citenamefont {Bethe}\ and\ \citenamefont
  {Bacher}(1936)}]{RevModPhys882}%
  \BibitemOpen
  \bibfield  {author} {\bibinfo {author} {\bibfnamefont {H.~A.}\ \bibnamefont
  {Bethe}}\ and\ \bibinfo {author} {\bibfnamefont {R.~F.}\ \bibnamefont
  {Bacher}},\ }\href {\doibase 10.1103/RevModPhys.8.82} {\bibfield  {journal}
  {\bibinfo  {journal} {Rev. Mod. Phys.}\ }\textbf {\bibinfo {volume} {8}},\
  \bibinfo {pages} {82} (\bibinfo {year} {1936})}\BibitemShut {NoStop}%
\bibitem [{\citenamefont {Möller}\ \emph {et~al.}(2016)\citenamefont
  {Möller}, \citenamefont {Sierk}, \citenamefont {Ichikawa},\ and\
  \citenamefont {Sagawa}}]{Moller1995}%
  \BibitemOpen
  \bibfield  {author} {\bibinfo {author} {\bibfnamefont {P.}~\bibnamefont
  {Möller}}, \bibinfo {author} {\bibfnamefont {A.}~\bibnamefont {Sierk}},
  \bibinfo {author} {\bibfnamefont {T.}~\bibnamefont {Ichikawa}}, \ and\
  \bibinfo {author} {\bibfnamefont {H.}~\bibnamefont {Sagawa}},\ }\href
  {\doibase https://doi.org/10.1016/j.adt.2015.10.002} {\bibfield  {journal}
  {\bibinfo  {journal} {At. Data Nucl. Data Tables}\ }\textbf {\bibinfo
  {volume} {109-110}},\ \bibinfo {pages} {1} (\bibinfo {year}
  {2016})}\BibitemShut {NoStop}%
\bibitem [{\citenamefont {Bethe}(1937)}]{RevModPhys969}%
  \BibitemOpen
  \bibfield  {author} {\bibinfo {author} {\bibfnamefont {H.~A.}\ \bibnamefont
  {Bethe}},\ }\href {\doibase 10.1103/RevModPhys.9.69} {\bibfield  {journal}
  {\bibinfo  {journal} {Rev. Mod. Phys.}\ }\textbf {\bibinfo {volume} {9}},\
  \bibinfo {pages} {69} (\bibinfo {year} {1937})}\BibitemShut {NoStop}%
\bibitem [{\citenamefont {Yalçın}(2017)}]{Yal2017}%
  \BibitemOpen
  \bibfield  {author} {\bibinfo {author} {\bibfnamefont {C.}~\bibnamefont
  {Yalçın}},\ }\href {\doibase https://doi.org/10.1007/s41365-017-0267-y}
  {\bibfield  {journal} {\bibinfo  {journal} {Nucl. Sci. Tech}\ }\textbf
  {\bibinfo {volume} {28}},\ \bibinfo {pages} {113} (\bibinfo {year}
  {2017})}\BibitemShut {NoStop}%
\bibitem [{\citenamefont {Luo}\ \emph {et~al.}(2023)\citenamefont {Luo},
  \citenamefont {Liang}, \citenamefont {Jiang}, \citenamefont {Zhou},\ and\
  \citenamefont {He}}]{Luo2023}%
  \BibitemOpen
  \bibfield  {author} {\bibinfo {author} {\bibfnamefont {J.-H.}\ \bibnamefont
  {Luo}}, \bibinfo {author} {\bibfnamefont {J.-C.}\ \bibnamefont {Liang}},
  \bibinfo {author} {\bibfnamefont {L.}~\bibnamefont {Jiang}}, \bibinfo
  {author} {\bibfnamefont {L.}~\bibnamefont {Zhou}}, \ and\ \bibinfo {author}
  {\bibfnamefont {L.}~\bibnamefont {He}},\ }\href {\doibase
  https://doi.org/10.1007/s41365-022-01158-z} {\bibfield  {journal} {\bibinfo
  {journal} {Nucl. Sci. Tech}\ }\textbf {\bibinfo {volume} {34}},\ \bibinfo
  {pages} {4} (\bibinfo {year} {2023})}\BibitemShut {NoStop}%
\bibitem [{\citenamefont {Chen}\ \emph
  {et~al.}(2023{\natexlab{a}})\citenamefont {Chen}, \citenamefont {Wu},
  \citenamefont {Yang}, \citenamefont {Zeng},\ and\ \citenamefont
  {Feng}}]{Chen2023}%
  \BibitemOpen
  \bibfield  {author} {\bibinfo {author} {\bibfnamefont {P.-H.}\ \bibnamefont
  {Chen}}, \bibinfo {author} {\bibfnamefont {H.}~\bibnamefont {Wu}}, \bibinfo
  {author} {\bibfnamefont {Z.-X.}\ \bibnamefont {Yang}}, \bibinfo {author}
  {\bibfnamefont {X.-H.}\ \bibnamefont {Zeng}}, \ and\ \bibinfo {author}
  {\bibfnamefont {Z.-Q.}\ \bibnamefont {Feng}},\ }\href {\doibase
  https://doi.org/10.1007/s41365-022-01157-0} {\bibfield  {journal} {\bibinfo
  {journal} {Nucl. Sci. Tech}\ }\textbf {\bibinfo {volume} {34}},\ \bibinfo
  {pages} {7} (\bibinfo {year} {2023}{\natexlab{a}})}\BibitemShut {NoStop}%
\bibitem [{\citenamefont {Bethe}(1936{\natexlab{a}})}]{PhysRev50332}%
  \BibitemOpen
  \bibfield  {author} {\bibinfo {author} {\bibfnamefont {H.~A.}\ \bibnamefont
  {Bethe}},\ }\href {\doibase 10.1103/PhysRev.50.332} {\bibfield  {journal}
  {\bibinfo  {journal} {Phys. Rev.}\ }\textbf {\bibinfo {volume} {50}},\
  \bibinfo {pages} {332} (\bibinfo {year} {1936}{\natexlab{a}})}\BibitemShut
  {NoStop}%
\bibitem [{\citenamefont {Dilg}\ \emph {et~al.}(1973)\citenamefont {Dilg},
  \citenamefont {Schantl}, \citenamefont {Vonach},\ and\ \citenamefont
  {Uhl}}]{DILG1973269}%
  \BibitemOpen
  \bibfield  {author} {\bibinfo {author} {\bibfnamefont {W.}~\bibnamefont
  {Dilg}}, \bibinfo {author} {\bibfnamefont {W.}~\bibnamefont {Schantl}},
  \bibinfo {author} {\bibfnamefont {H.}~\bibnamefont {Vonach}}, \ and\ \bibinfo
  {author} {\bibfnamefont {M.}~\bibnamefont {Uhl}},\ }\href {\doibase
  https://doi.org/10.1016/0375-9474(73)90196-6} {\bibfield  {journal} {\bibinfo
   {journal} {Nucl. Phys. A}\ }\textbf {\bibinfo {volume} {217}},\ \bibinfo
  {pages} {269} (\bibinfo {year} {1973})}\BibitemShut {NoStop}%
\bibitem [{\citenamefont {Gilbert}\ and\ \citenamefont
  {Cameron}(1965)}]{doi101139p65139}%
  \BibitemOpen
  \bibfield  {author} {\bibinfo {author} {\bibfnamefont {A.}~\bibnamefont
  {Gilbert}}\ and\ \bibinfo {author} {\bibfnamefont {A.~G.~W.}\ \bibnamefont
  {Cameron}},\ }\href {\doibase 10.1139/p65-139} {\bibfield  {journal}
  {\bibinfo  {journal} {Canadian Journal of Physics}\ }\textbf {\bibinfo
  {volume} {43}},\ \bibinfo {pages} {1446} (\bibinfo {year}
  {1965})}\BibitemShut {NoStop}%
\bibitem [{\citenamefont {Koning}\ \emph {et~al.}(2008)\citenamefont {Koning},
  \citenamefont {Hilaire},\ and\ \citenamefont {Goriely}}]{KONING200813}%
  \BibitemOpen
  \bibfield  {author} {\bibinfo {author} {\bibfnamefont {A.}~\bibnamefont
  {Koning}}, \bibinfo {author} {\bibfnamefont {S.}~\bibnamefont {Hilaire}}, \
  and\ \bibinfo {author} {\bibfnamefont {S.}~\bibnamefont {Goriely}},\ }\href
  {\doibase https://doi.org/10.1016/j.nuclphysa.2008.06.005} {\bibfield
  {journal} {\bibinfo  {journal} {Nucl. Phys. A}\ }\textbf {\bibinfo {volume}
  {810}},\ \bibinfo {pages} {13} (\bibinfo {year} {2008})}\BibitemShut
  {NoStop}%
\bibitem [{\citenamefont {Hilaire}\ \emph {et~al.}(2001)\citenamefont
  {Hilaire}, \citenamefont {Delaroche},\ and\ \citenamefont {Girod}}]{3}%
  \BibitemOpen
  \bibfield  {author} {\bibinfo {author} {\bibfnamefont {S.}~\bibnamefont
  {Hilaire}}, \bibinfo {author} {\bibfnamefont {J.}~\bibnamefont {Delaroche}},
  \ and\ \bibinfo {author} {\bibfnamefont {M.}~\bibnamefont {Girod}},\ }\href
  {\doibase 10.1007/s100500170025} {\bibfield  {journal} {\bibinfo  {journal}
  {Eur. Phys. J A}\ }\textbf {\bibinfo {volume} {12}},\ \bibinfo {pages} {169}
  (\bibinfo {year} {2001})}\BibitemShut {NoStop}%
\bibitem [{\citenamefont {Williams}(1971)}]{WILLIAMS1971231}%
  \BibitemOpen
  \bibfield  {author} {\bibinfo {author} {\bibfnamefont {F.~C.}\ \bibnamefont
  {Williams}},\ }\href {\doibase https://doi.org/10.1016/0375-9474(71)90426-X}
  {\bibfield  {journal} {\bibinfo  {journal} {Nucl. Phys. A}\ }\textbf
  {\bibinfo {volume} {166}},\ \bibinfo {pages} {231} (\bibinfo {year}
  {1971})}\BibitemShut {NoStop}%
\bibitem [{\citenamefont {Běták}\ and\ \citenamefont {Dobeš}(1976)}]{19712}%
  \BibitemOpen
  \bibfield  {author} {\bibinfo {author} {\bibfnamefont {E.}~\bibnamefont
  {Běták}}\ and\ \bibinfo {author} {\bibfnamefont {J.}~\bibnamefont
  {Dobeš}},\ }\href {\doibase https://doi.org/10.1007/BF01408305} {\bibfield
  {journal} {\bibinfo  {journal} {Z. Phys. A}\ }\textbf {\bibinfo {volume}
  {279}},\ \bibinfo {pages} {319} (\bibinfo {year} {1976})}\BibitemShut
  {NoStop}%
\bibitem [{\citenamefont {Obložinský}(1986)}]{OBLOZINSKY1986127}%
  \BibitemOpen
  \bibfield  {author} {\bibinfo {author} {\bibfnamefont {P.}~\bibnamefont
  {Obložinský}},\ }\href {\doibase
  https://doi.org/10.1016/0375-9474(86)90033-3} {\bibfield  {journal} {\bibinfo
   {journal} {Nucl. Phys. A}\ }\textbf {\bibinfo {volume} {453}},\ \bibinfo
  {pages} {127} (\bibinfo {year} {1986})}\BibitemShut {NoStop}%
\bibitem [{\citenamefont {Hilaire}\ \emph {et~al.}(1998)\citenamefont
  {Hilaire}, \citenamefont {Delaroche},\ and\ \citenamefont
  {Koning}}]{HILAIRE1998417}%
  \BibitemOpen
  \bibfield  {author} {\bibinfo {author} {\bibfnamefont {S.}~\bibnamefont
  {Hilaire}}, \bibinfo {author} {\bibfnamefont {J.}~\bibnamefont {Delaroche}},
  \ and\ \bibinfo {author} {\bibfnamefont {A.}~\bibnamefont {Koning}},\ }\href
  {\doibase https://doi.org/10.1016/S0375-9474(98)00003-7} {\bibfield
  {journal} {\bibinfo  {journal} {Nucl. Phys. A}\ }\textbf {\bibinfo {volume}
  {632}},\ \bibinfo {pages} {417} (\bibinfo {year} {1998})}\BibitemShut
  {NoStop}%
\bibitem [{\citenamefont {Alhassid}\ \emph {et~al.}(1999)\citenamefont
  {Alhassid}, \citenamefont {Liu},\ and\ \citenamefont
  {Nakada}}]{PhysRevLett834265}%
  \BibitemOpen
  \bibfield  {author} {\bibinfo {author} {\bibfnamefont {Y.}~\bibnamefont
  {Alhassid}}, \bibinfo {author} {\bibfnamefont {S.}~\bibnamefont {Liu}}, \
  and\ \bibinfo {author} {\bibfnamefont {H.}~\bibnamefont {Nakada}},\ }\href
  {\doibase 10.1103/PhysRevLett.83.4265} {\bibfield  {journal} {\bibinfo
  {journal} {Phys. Rev. Lett.}\ }\textbf {\bibinfo {volume} {83}},\ \bibinfo
  {pages} {4265} (\bibinfo {year} {1999})}\BibitemShut {NoStop}%
\bibitem [{\citenamefont {Ormand}(1997)}]{PhysRevC56R1678}%
  \BibitemOpen
  \bibfield  {author} {\bibinfo {author} {\bibfnamefont {W.~E.}\ \bibnamefont
  {Ormand}},\ }\href {\doibase 10.1103/PhysRevC.56.R1678} {\bibfield  {journal}
  {\bibinfo  {journal} {Phys. Rev. C}\ }\textbf {\bibinfo {volume} {56}},\
  \bibinfo {pages} {R1678} (\bibinfo {year} {1997})}\BibitemShut {NoStop}%
\bibitem [{\citenamefont {White}\ \emph {et~al.}(2000)\citenamefont {White},
  \citenamefont {Koonin},\ and\ \citenamefont {Dean}}]{PhysRevC61034303}%
  \BibitemOpen
  \bibfield  {author} {\bibinfo {author} {\bibfnamefont {J.~A.}\ \bibnamefont
  {White}}, \bibinfo {author} {\bibfnamefont {S.~E.}\ \bibnamefont {Koonin}}, \
  and\ \bibinfo {author} {\bibfnamefont {D.~J.}\ \bibnamefont {Dean}},\ }\href
  {\doibase 10.1103/PhysRevC.61.034303} {\bibfield  {journal} {\bibinfo
  {journal} {Phys. Rev. C}\ }\textbf {\bibinfo {volume} {61}},\ \bibinfo
  {pages} {034303} (\bibinfo {year} {2000})}\BibitemShut {NoStop}%
\bibitem [{\citenamefont {Cerf}(1994{\natexlab{a}})}]{PhysRevC49852}%
  \BibitemOpen
  \bibfield  {author} {\bibinfo {author} {\bibfnamefont {N.}~\bibnamefont
  {Cerf}},\ }\href {\doibase 10.1103/PhysRevC.49.852} {\bibfield  {journal}
  {\bibinfo  {journal} {Phys. Rev. C}\ }\textbf {\bibinfo {volume} {49}},\
  \bibinfo {pages} {852} (\bibinfo {year} {1994}{\natexlab{a}})}\BibitemShut
  {NoStop}%
\bibitem [{\citenamefont {Cerf}(1994{\natexlab{b}})}]{PhysRevC50836}%
  \BibitemOpen
  \bibfield  {author} {\bibinfo {author} {\bibfnamefont {N.}~\bibnamefont
  {Cerf}},\ }\href {\doibase 10.1103/PhysRevC.50.836} {\bibfield  {journal}
  {\bibinfo  {journal} {Phys. Rev. C}\ }\textbf {\bibinfo {volume} {50}},\
  \bibinfo {pages} {836} (\bibinfo {year} {1994}{\natexlab{b}})}\BibitemShut
  {NoStop}%
\bibitem [{\citenamefont {Strohmaier}\ \emph {et~al.}(1987)\citenamefont
  {Strohmaier}, \citenamefont {Grimes},\ and\ \citenamefont
  {Satyanarayana}}]{PhysRevC361604}%
  \BibitemOpen
  \bibfield  {author} {\bibinfo {author} {\bibfnamefont {B.}~\bibnamefont
  {Strohmaier}}, \bibinfo {author} {\bibfnamefont {S.~M.}\ \bibnamefont
  {Grimes}}, \ and\ \bibinfo {author} {\bibfnamefont {H.}~\bibnamefont
  {Satyanarayana}},\ }\href {\doibase 10.1103/PhysRevC.36.1604} {\bibfield
  {journal} {\bibinfo  {journal} {Phys. Rev. C}\ }\textbf {\bibinfo {volume}
  {36}},\ \bibinfo {pages} {1604} (\bibinfo {year} {1987})}\BibitemShut
  {NoStop}%
\bibitem [{\citenamefont {Grimes}\ and\ \citenamefont
  {Massey}(1995)}]{PhysRevC51606}%
  \BibitemOpen
  \bibfield  {author} {\bibinfo {author} {\bibfnamefont {S.~M.}\ \bibnamefont
  {Grimes}}\ and\ \bibinfo {author} {\bibfnamefont {T.~N.}\ \bibnamefont
  {Massey}},\ }\href {\doibase 10.1103/PhysRevC.51.606} {\bibfield  {journal}
  {\bibinfo  {journal} {Phys. Rev. C}\ }\textbf {\bibinfo {volume} {51}},\
  \bibinfo {pages} {606} (\bibinfo {year} {1995})}\BibitemShut {NoStop}%
\bibitem [{\citenamefont {French}\ and\ \citenamefont
  {Ratcliff}(1971)}]{PhysRevC394}%
  \BibitemOpen
  \bibfield  {author} {\bibinfo {author} {\bibfnamefont {J.~B.}\ \bibnamefont
  {French}}\ and\ \bibinfo {author} {\bibfnamefont {K.~F.}\ \bibnamefont
  {Ratcliff}},\ }\href {\doibase 10.1103/PhysRevC.3.94} {\bibfield  {journal}
  {\bibinfo  {journal} {Phys. Rev. C}\ }\textbf {\bibinfo {volume} {3}},\
  \bibinfo {pages} {94} (\bibinfo {year} {1971})}\BibitemShut {NoStop}%
\bibitem [{\citenamefont {Dang}\ \emph {et~al.}(2017)\citenamefont {Dang},
  \citenamefont {Hung},\ and\ \citenamefont {Huong}}]{PhysRevC.96.054321}%
  \BibitemOpen
  \bibfield  {author} {\bibinfo {author} {\bibfnamefont {N.~D.}\ \bibnamefont
  {Dang}}, \bibinfo {author} {\bibfnamefont {N.~Q.}\ \bibnamefont {Hung}}, \
  and\ \bibinfo {author} {\bibfnamefont {L.~T.~Q.}\ \bibnamefont {Huong}},\
  }\href {\doibase 10.1103/PhysRevC.96.054321} {\bibfield  {journal} {\bibinfo
  {journal} {Phys. Rev. C}\ }\textbf {\bibinfo {volume} {96}},\ \bibinfo
  {pages} {054321} (\bibinfo {year} {2017})}\BibitemShut {NoStop}%
\bibitem [{\citenamefont {Hung}\ \emph {et~al.}(2017)\citenamefont {Hung},
  \citenamefont {Dang},\ and\ \citenamefont {Huong}}]{PhysRevLett.118.022502}%
  \BibitemOpen
  \bibfield  {author} {\bibinfo {author} {\bibfnamefont {N.~Q.}\ \bibnamefont
  {Hung}}, \bibinfo {author} {\bibfnamefont {N.~D.}\ \bibnamefont {Dang}}, \
  and\ \bibinfo {author} {\bibfnamefont {L.~T.~Q.}\ \bibnamefont {Huong}},\
  }\href {\doibase 10.1103/PhysRevLett.118.022502} {\bibfield  {journal}
  {\bibinfo  {journal} {Phys. Rev. Lett.}\ }\textbf {\bibinfo {volume} {118}},\
  \bibinfo {pages} {022502} (\bibinfo {year} {2017})}\BibitemShut {NoStop}%
\bibitem [{\citenamefont {Dey}\ \emph {et~al.}(2017)\citenamefont {Dey},
  \citenamefont {Pandit}, \citenamefont {Bhattacharya}, \citenamefont {Hung},
  \citenamefont {Dang}, \citenamefont {Phuc}, \citenamefont {Mondal},
  \citenamefont {Mukhopadhyay}, \citenamefont {Pal}, \citenamefont {De},\ and\
  \citenamefont {Banerjee}}]{PhysRevC.96.054326}%
  \BibitemOpen
  \bibfield  {author} {\bibinfo {author} {\bibfnamefont {B.}~\bibnamefont
  {Dey}}, \bibinfo {author} {\bibfnamefont {D.}~\bibnamefont {Pandit}},
  \bibinfo {author} {\bibfnamefont {S.}~\bibnamefont {Bhattacharya}}, \bibinfo
  {author} {\bibfnamefont {N.~Q.}\ \bibnamefont {Hung}}, \bibinfo {author}
  {\bibfnamefont {N.~D.}\ \bibnamefont {Dang}}, \bibinfo {author}
  {\bibfnamefont {L.~T.}\ \bibnamefont {Phuc}}, \bibinfo {author}
  {\bibfnamefont {D.}~\bibnamefont {Mondal}}, \bibinfo {author} {\bibfnamefont
  {S.}~\bibnamefont {Mukhopadhyay}}, \bibinfo {author} {\bibfnamefont
  {S.}~\bibnamefont {Pal}}, \bibinfo {author} {\bibfnamefont {A.}~\bibnamefont
  {De}}, \ and\ \bibinfo {author} {\bibfnamefont {S.~R.}\ \bibnamefont
  {Banerjee}},\ }\href {\doibase 10.1103/PhysRevC.96.054326} {\bibfield
  {journal} {\bibinfo  {journal} {Phys. Rev. C}\ }\textbf {\bibinfo {volume}
  {96}},\ \bibinfo {pages} {054326} (\bibinfo {year} {2017})}\BibitemShut
  {NoStop}%
\bibitem [{\citenamefont {Dey}\ \emph {et~al.}(2019)\citenamefont {Dey},
  \citenamefont {{Quang Hung}}, \citenamefont {Pandit}, \citenamefont
  {Bhattacharya}, \citenamefont {{Dinh Dang}}, \citenamefont {{Quynh Huong}},
  \citenamefont {Mondal}, \citenamefont {Mukhopadhyay}, \citenamefont {Pal},
  \citenamefont {De},\ and\ \citenamefont {Banerjee}}]{DEY2019634}%
  \BibitemOpen
  \bibfield  {author} {\bibinfo {author} {\bibfnamefont {B.}~\bibnamefont
  {Dey}}, \bibinfo {author} {\bibfnamefont {N.}~\bibnamefont {{Quang Hung}}},
  \bibinfo {author} {\bibfnamefont {D.}~\bibnamefont {Pandit}}, \bibinfo
  {author} {\bibfnamefont {S.}~\bibnamefont {Bhattacharya}}, \bibinfo {author}
  {\bibfnamefont {N.}~\bibnamefont {{Dinh Dang}}}, \bibinfo {author}
  {\bibfnamefont {L.}~\bibnamefont {{Quynh Huong}}}, \bibinfo {author}
  {\bibfnamefont {D.}~\bibnamefont {Mondal}}, \bibinfo {author} {\bibfnamefont
  {S.}~\bibnamefont {Mukhopadhyay}}, \bibinfo {author} {\bibfnamefont
  {S.}~\bibnamefont {Pal}}, \bibinfo {author} {\bibfnamefont {A.}~\bibnamefont
  {De}}, \ and\ \bibinfo {author} {\bibfnamefont {S.}~\bibnamefont
  {Banerjee}},\ }\href {\doibase
  https://doi.org/10.1016/j.physletb.2018.12.007} {\bibfield  {journal}
  {\bibinfo  {journal} {Phys. Lett. B}\ }\textbf {\bibinfo {volume} {789}},\
  \bibinfo {pages} {634} (\bibinfo {year} {2019})}\BibitemShut {NoStop}%
\bibitem [{\citenamefont {Agrawal}\ and\ \citenamefont
  {Ansari}(1998)}]{AGRAWAL1998362}%
  \BibitemOpen
  \bibfield  {author} {\bibinfo {author} {\bibfnamefont {B.}~\bibnamefont
  {Agrawal}}\ and\ \bibinfo {author} {\bibfnamefont {A.}~\bibnamefont
  {Ansari}},\ }\href {\doibase https://doi.org/10.1016/S0375-9474(98)00462-X}
  {\bibfield  {journal} {\bibinfo  {journal} {Nucl. Phys. A}\ }\textbf
  {\bibinfo {volume} {640}},\ \bibinfo {pages} {362} (\bibinfo {year}
  {1998})}\BibitemShut {NoStop}%
\bibitem [{\citenamefont {Goriely}(1996)}]{GORIELY199628}%
  \BibitemOpen
  \bibfield  {author} {\bibinfo {author} {\bibfnamefont {S.}~\bibnamefont
  {Goriely}},\ }\href {\doibase https://doi.org/10.1016/0375-9474(96)00162-5}
  {\bibfield  {journal} {\bibinfo  {journal} {Nucl. Phys. A}\ }\textbf
  {\bibinfo {volume} {605}},\ \bibinfo {pages} {28} (\bibinfo {year}
  {1996})}\BibitemShut {NoStop}%
\bibitem [{\citenamefont {Decowski}\ \emph {et~al.}(1968)\citenamefont
  {Decowski}, \citenamefont {Grochulski}, \citenamefont {Marcinkowski},
  \citenamefont {Siwek},\ and\ \citenamefont {Wilhelmi}}]{DECOWSKI1968129}%
  \BibitemOpen
  \bibfield  {author} {\bibinfo {author} {\bibfnamefont {P.}~\bibnamefont
  {Decowski}}, \bibinfo {author} {\bibfnamefont {W.}~\bibnamefont
  {Grochulski}}, \bibinfo {author} {\bibfnamefont {A.}~\bibnamefont
  {Marcinkowski}}, \bibinfo {author} {\bibfnamefont {K.}~\bibnamefont {Siwek}},
  \ and\ \bibinfo {author} {\bibfnamefont {Z.}~\bibnamefont {Wilhelmi}},\
  }\href {\doibase https://doi.org/10.1016/0375-9474(68)90687-8} {\bibfield
  {journal} {\bibinfo  {journal} {Nucl. Phys. A}\ }\textbf {\bibinfo {volume}
  {110}},\ \bibinfo {pages} {129} (\bibinfo {year} {1968})}\BibitemShut
  {NoStop}%
\bibitem [{\citenamefont {Demetriou}\ and\ \citenamefont
  {Goriely}(2001)}]{DEMETRIOU200195}%
  \BibitemOpen
  \bibfield  {author} {\bibinfo {author} {\bibfnamefont {P.}~\bibnamefont
  {Demetriou}}\ and\ \bibinfo {author} {\bibfnamefont {S.}~\bibnamefont
  {Goriely}},\ }\href {\doibase https://doi.org/10.1016/S0375-9474(01)01095-8}
  {\bibfield  {journal} {\bibinfo  {journal} {Nucl. Phys. A}\ }\textbf
  {\bibinfo {volume} {695}},\ \bibinfo {pages} {95} (\bibinfo {year}
  {2001})}\BibitemShut {NoStop}%
\bibitem [{\citenamefont {Papenbrock}\ and\ \citenamefont
  {Weidenm\"uller}(2007)}]{RevModPhys.79.997}%
  \BibitemOpen
  \bibfield  {author} {\bibinfo {author} {\bibfnamefont {T.}~\bibnamefont
  {Papenbrock}}\ and\ \bibinfo {author} {\bibfnamefont {H.~A.}\ \bibnamefont
  {Weidenm\"uller}},\ }\href {\doibase 10.1103/RevModPhys.79.997} {\bibfield
  {journal} {\bibinfo  {journal} {Rev. Mod. Phys.}\ }\textbf {\bibinfo {volume}
  {79}},\ \bibinfo {pages} {997} (\bibinfo {year} {2007})}\BibitemShut
  {NoStop}%
\bibitem [{\citenamefont {Shimizu}\ \emph {et~al.}(2016)\citenamefont
  {Shimizu}, \citenamefont {Utsuno}, \citenamefont {Futamura}, \citenamefont
  {Sakurai}, \citenamefont {Mizusaki},\ and\ \citenamefont
  {Otsuka}}]{SHIMIZU201613}%
  \BibitemOpen
  \bibfield  {author} {\bibinfo {author} {\bibfnamefont {N.}~\bibnamefont
  {Shimizu}}, \bibinfo {author} {\bibfnamefont {Y.}~\bibnamefont {Utsuno}},
  \bibinfo {author} {\bibfnamefont {Y.}~\bibnamefont {Futamura}}, \bibinfo
  {author} {\bibfnamefont {T.}~\bibnamefont {Sakurai}}, \bibinfo {author}
  {\bibfnamefont {T.}~\bibnamefont {Mizusaki}}, \ and\ \bibinfo {author}
  {\bibfnamefont {T.}~\bibnamefont {Otsuka}},\ }\href {\doibase
  https://doi.org/10.1016/j.physletb.2015.12.005} {\bibfield  {journal}
  {\bibinfo  {journal} {Phys. Lett. B}\ }\textbf {\bibinfo {volume} {753}},\
  \bibinfo {pages} {13} (\bibinfo {year} {2016})}\BibitemShut {NoStop}%
\bibitem [{\citenamefont {Chen}\ \emph
  {et~al.}(2023{\natexlab{b}})\citenamefont {Chen}, \citenamefont {Liu},
  \citenamefont {Yuan}, \citenamefont {Chen}, \citenamefont {Shimizu},
  \citenamefont {Sun}, \citenamefont {Xu},\ and\ \citenamefont
  {Tian}}]{PhysRevC.107.054306}%
  \BibitemOpen
  \bibfield  {author} {\bibinfo {author} {\bibfnamefont {J.}~\bibnamefont
  {Chen}}, \bibinfo {author} {\bibfnamefont {M.}~\bibnamefont {Liu}}, \bibinfo
  {author} {\bibfnamefont {C.}~\bibnamefont {Yuan}}, \bibinfo {author}
  {\bibfnamefont {S.}~\bibnamefont {Chen}}, \bibinfo {author} {\bibfnamefont
  {N.}~\bibnamefont {Shimizu}}, \bibinfo {author} {\bibfnamefont
  {X.}~\bibnamefont {Sun}}, \bibinfo {author} {\bibfnamefont {R.}~\bibnamefont
  {Xu}}, \ and\ \bibinfo {author} {\bibfnamefont {Y.}~\bibnamefont {Tian}},\
  }\href {\doibase 10.1103/PhysRevC.107.054306} {\bibfield  {journal} {\bibinfo
   {journal} {Phys. Rev. C}\ }\textbf {\bibinfo {volume} {107}},\ \bibinfo
  {pages} {054306} (\bibinfo {year} {2023}{\natexlab{b}})}\BibitemShut
  {NoStop}%
\bibitem [{\citenamefont {Berger}\ and\ \citenamefont
  {Martinot}(1974)}]{BERGER1974391}%
  \BibitemOpen
  \bibfield  {author} {\bibinfo {author} {\bibfnamefont {J.}~\bibnamefont
  {Berger}}\ and\ \bibinfo {author} {\bibfnamefont {M.}~\bibnamefont
  {Martinot}},\ }\href {\doibase https://doi.org/10.1016/0375-9474(74)90491-6}
  {\bibfield  {journal} {\bibinfo  {journal} {Nucl Phys A}\ }\textbf {\bibinfo
  {volume} {226}},\ \bibinfo {pages} {391} (\bibinfo {year}
  {1974})}\BibitemShut {NoStop}%
\bibitem [{\citenamefont {Girod}\ \emph {et~al.}(1988)\citenamefont {Girod},
  \citenamefont {Dessagne}, \citenamefont {Bernas}, \citenamefont {Langevin},
  \citenamefont {Pougheon},\ and\ \citenamefont {Roussel}}]{PhysRevC.37.2600}%
  \BibitemOpen
  \bibfield  {author} {\bibinfo {author} {\bibfnamefont {M.}~\bibnamefont
  {Girod}}, \bibinfo {author} {\bibfnamefont {P.}~\bibnamefont {Dessagne}},
  \bibinfo {author} {\bibfnamefont {M.}~\bibnamefont {Bernas}}, \bibinfo
  {author} {\bibfnamefont {M.}~\bibnamefont {Langevin}}, \bibinfo {author}
  {\bibfnamefont {F.}~\bibnamefont {Pougheon}}, \ and\ \bibinfo {author}
  {\bibfnamefont {P.}~\bibnamefont {Roussel}},\ }\href {\doibase
  10.1103/PhysRevC.37.2600} {\bibfield  {journal} {\bibinfo  {journal} {Phys.
  Rev. C}\ }\textbf {\bibinfo {volume} {37}},\ \bibinfo {pages} {2600}
  (\bibinfo {year} {1988})}\BibitemShut {NoStop}%
\bibitem [{\citenamefont {Goriely}\ \emph {et~al.}(2008)\citenamefont
  {Goriely}, \citenamefont {Hilaire},\ and\ \citenamefont
  {Koning}}]{PhysRevC.78.064307}%
  \BibitemOpen
  \bibfield  {author} {\bibinfo {author} {\bibfnamefont {S.}~\bibnamefont
  {Goriely}}, \bibinfo {author} {\bibfnamefont {S.}~\bibnamefont {Hilaire}}, \
  and\ \bibinfo {author} {\bibfnamefont {A.~J.}\ \bibnamefont {Koning}},\
  }\href {\doibase 10.1103/PhysRevC.78.064307} {\bibfield  {journal} {\bibinfo
  {journal} {Phys. Rev. C}\ }\textbf {\bibinfo {volume} {78}},\ \bibinfo
  {pages} {064307} (\bibinfo {year} {2008})}\BibitemShut {NoStop}%
\bibitem [{\citenamefont {Hilaire}\ \emph {et~al.}(2012)\citenamefont
  {Hilaire}, \citenamefont {Girod}, \citenamefont {Goriely},\ and\
  \citenamefont {Koning}}]{PhysRevC.86.064317}%
  \BibitemOpen
  \bibfield  {author} {\bibinfo {author} {\bibfnamefont {S.}~\bibnamefont
  {Hilaire}}, \bibinfo {author} {\bibfnamefont {M.}~\bibnamefont {Girod}},
  \bibinfo {author} {\bibfnamefont {S.}~\bibnamefont {Goriely}}, \ and\
  \bibinfo {author} {\bibfnamefont {A.~J.}\ \bibnamefont {Koning}},\ }\href
  {\doibase 10.1103/PhysRevC.86.064317} {\bibfield  {journal} {\bibinfo
  {journal} {Phys. Rev. C}\ }\textbf {\bibinfo {volume} {86}},\ \bibinfo
  {pages} {064317} (\bibinfo {year} {2012})}\BibitemShut {NoStop}%
\bibitem [{\citenamefont {Ring}(1996)}]{RING1996193}%
  \BibitemOpen
  \bibfield  {author} {\bibinfo {author} {\bibfnamefont {P.}~\bibnamefont
  {Ring}},\ }\href {\doibase https://doi.org/10.1016/0146-6410(96)00054-3}
  {\bibfield  {journal} {\bibinfo  {journal} {Prog. Part. Nucl. Phys.}\
  }\textbf {\bibinfo {volume} {37}},\ \bibinfo {pages} {193} (\bibinfo {year}
  {1996})}\BibitemShut {NoStop}%
\bibitem [{\citenamefont {Meng}\ \emph {et~al.}(2006)\citenamefont {Meng},
  \citenamefont {Toki}, \citenamefont {Zhou}, \citenamefont {Zhang},
  \citenamefont {Long},\ and\ \citenamefont {Geng}}]{MENG2006470}%
  \BibitemOpen
  \bibfield  {author} {\bibinfo {author} {\bibfnamefont {J.}~\bibnamefont
  {Meng}}, \bibinfo {author} {\bibfnamefont {H.}~\bibnamefont {Toki}}, \bibinfo
  {author} {\bibfnamefont {S.}~\bibnamefont {Zhou}}, \bibinfo {author}
  {\bibfnamefont {S.}~\bibnamefont {Zhang}}, \bibinfo {author} {\bibfnamefont
  {W.}~\bibnamefont {Long}}, \ and\ \bibinfo {author} {\bibfnamefont
  {L.}~\bibnamefont {Geng}},\ }\href {\doibase
  https://doi.org/10.1016/j.ppnp.2005.06.001} {\bibfield  {journal} {\bibinfo
  {journal} {Prog. Part. Nucl. Phys.}\ }\textbf {\bibinfo {volume} {57}},\
  \bibinfo {pages} {470} (\bibinfo {year} {2006})}\BibitemShut {NoStop}%
\bibitem [{\citenamefont {Vretenar}\ \emph {et~al.}(2005)\citenamefont
  {Vretenar}, \citenamefont {Afanasjev}, \citenamefont {Lalazissis},\ and\
  \citenamefont {Ring}}]{VRETENAR2005101}%
  \BibitemOpen
  \bibfield  {author} {\bibinfo {author} {\bibfnamefont {D.}~\bibnamefont
  {Vretenar}}, \bibinfo {author} {\bibfnamefont {A.}~\bibnamefont {Afanasjev}},
  \bibinfo {author} {\bibfnamefont {G.}~\bibnamefont {Lalazissis}}, \ and\
  \bibinfo {author} {\bibfnamefont {P.}~\bibnamefont {Ring}},\ }\href {\doibase
  https://doi.org/10.1016/j.physrep.2004.10.001} {\bibfield  {journal}
  {\bibinfo  {journal} {Phys. Rep.}\ }\textbf {\bibinfo {volume} {409}},\
  \bibinfo {pages} {101} (\bibinfo {year} {2005})}\BibitemShut {NoStop}%
\bibitem [{\citenamefont {Nikšić}\ \emph {et~al.}(2011)\citenamefont
  {Nikšić}, \citenamefont {Vretenar},\ and\ \citenamefont
  {Ring}}]{NIKSIC2011519}%
  \BibitemOpen
  \bibfield  {author} {\bibinfo {author} {\bibfnamefont {T.}~\bibnamefont
  {Nikšić}}, \bibinfo {author} {\bibfnamefont {D.}~\bibnamefont {Vretenar}},
  \ and\ \bibinfo {author} {\bibfnamefont {P.}~\bibnamefont {Ring}},\ }\href
  {\doibase https://doi.org/10.1016/j.ppnp.2011.01.055} {\bibfield  {journal}
  {\bibinfo  {journal} {Prog. Part. Nucl. Phys.}\ }\textbf {\bibinfo {volume}
  {66}},\ \bibinfo {pages} {519} (\bibinfo {year} {2011})}\BibitemShut
  {NoStop}%
\bibitem [{\citenamefont {Meng}\ \emph {et~al.}(2013)\citenamefont {Meng},
  \citenamefont {Peng}, \citenamefont {Zhang},\ and\ \citenamefont
  {Zhao}}]{Meng2013}%
  \BibitemOpen
  \bibfield  {author} {\bibinfo {author} {\bibfnamefont {J.}~\bibnamefont
  {Meng}}, \bibinfo {author} {\bibfnamefont {J.}~\bibnamefont {Peng}}, \bibinfo
  {author} {\bibfnamefont {S.-Q.}\ \bibnamefont {Zhang}}, \ and\ \bibinfo
  {author} {\bibfnamefont {P.-W.}\ \bibnamefont {Zhao}},\ }\href {\doibase
  https://doi.org/10.1007/s11467-013-0287-y} {\bibfield  {journal} {\bibinfo
  {journal} {Front. Phys.}\ }\textbf {\bibinfo {volume} {8}},\ \bibinfo {pages}
  {55} (\bibinfo {year} {2013})}\BibitemShut {NoStop}%
\bibitem [{\citenamefont {Meng}(2015)}]{Meng2015}%
  \BibitemOpen
  \bibfield  {author} {\bibinfo {author} {\bibfnamefont {J.}~\bibnamefont
  {Meng}},\ }\href {\doibase 10.1142/9872} {\emph {\bibinfo {title}
  {Relativistic Density Functional for Nuclear Structure}}}\ (\bibinfo
  {publisher} {World Scientific, Singapore},\ \bibinfo {year}
  {2015})\BibitemShut {NoStop}%
\bibitem [{\citenamefont {Shen}\ \emph {et~al.}(2019)\citenamefont {Shen},
  \citenamefont {Liang}, \citenamefont {Long}, \citenamefont {Meng},\ and\
  \citenamefont {Ring}}]{SHEN2019103713}%
  \BibitemOpen
  \bibfield  {author} {\bibinfo {author} {\bibfnamefont {S.}~\bibnamefont
  {Shen}}, \bibinfo {author} {\bibfnamefont {H.}~\bibnamefont {Liang}},
  \bibinfo {author} {\bibfnamefont {W.~H.}\ \bibnamefont {Long}}, \bibinfo
  {author} {\bibfnamefont {J.}~\bibnamefont {Meng}}, \ and\ \bibinfo {author}
  {\bibfnamefont {P.}~\bibnamefont {Ring}},\ }\href {\doibase
  https://doi.org/10.1016/j.ppnp.2019.103713} {\bibfield  {journal} {\bibinfo
  {journal} {Prog. Part. Nucl. Phys}\ }\textbf {\bibinfo {volume} {109}},\
  \bibinfo {pages} {103713} (\bibinfo {year} {2019})}\BibitemShut {NoStop}%
\bibitem [{\citenamefont {Sharma}\ \emph {et~al.}(1993)\citenamefont {Sharma},
  \citenamefont {Lalazissis},\ and\ \citenamefont {Ring}}]{SHARMA19939}%
  \BibitemOpen
  \bibfield  {author} {\bibinfo {author} {\bibfnamefont {M.}~\bibnamefont
  {Sharma}}, \bibinfo {author} {\bibfnamefont {G.}~\bibnamefont {Lalazissis}},
  \ and\ \bibinfo {author} {\bibfnamefont {P.}~\bibnamefont {Ring}},\ }\href
  {\doibase https://doi.org/10.1016/0370-2693(93)91561-Z} {\bibfield  {journal}
  {\bibinfo  {journal} {Phys. Lett. B}\ }\textbf {\bibinfo {volume} {317}},\
  \bibinfo {pages} {9} (\bibinfo {year} {1993})}\BibitemShut {NoStop}%
\bibitem [{\citenamefont {Zhou}\ \emph {et~al.}(2003)\citenamefont {Zhou},
  \citenamefont {Meng},\ and\ \citenamefont {Ring}}]{PhysRevLett.91.262501}%
  \BibitemOpen
  \bibfield  {author} {\bibinfo {author} {\bibfnamefont {S.-G.}\ \bibnamefont
  {Zhou}}, \bibinfo {author} {\bibfnamefont {J.}~\bibnamefont {Meng}}, \ and\
  \bibinfo {author} {\bibfnamefont {P.}~\bibnamefont {Ring}},\ }\href {\doibase
  10.1103/PhysRevLett.91.262501} {\bibfield  {journal} {\bibinfo  {journal}
  {Phys. Rev. Lett.}\ }\textbf {\bibinfo {volume} {91}},\ \bibinfo {pages}
  {262501} (\bibinfo {year} {2003})}\BibitemShut {NoStop}%
\bibitem [{\citenamefont {Li}\ \emph {et~al.}(2013)\citenamefont {Li},
  \citenamefont {Wei}, \citenamefont {Hu}, \citenamefont {Ring},\ and\
  \citenamefont {Meng}}]{PhysRevC.88.064307}%
  \BibitemOpen
  \bibfield  {author} {\bibinfo {author} {\bibfnamefont {J.}~\bibnamefont
  {Li}}, \bibinfo {author} {\bibfnamefont {J.~X.}\ \bibnamefont {Wei}},
  \bibinfo {author} {\bibfnamefont {J.~N.}\ \bibnamefont {Hu}}, \bibinfo
  {author} {\bibfnamefont {P.}~\bibnamefont {Ring}}, \ and\ \bibinfo {author}
  {\bibfnamefont {J.}~\bibnamefont {Meng}},\ }\href {\doibase
  10.1103/PhysRevC.88.064307} {\bibfield  {journal} {\bibinfo  {journal} {Phys.
  Rev. C}\ }\textbf {\bibinfo {volume} {88}},\ \bibinfo {pages} {064307}
  (\bibinfo {year} {2013})}\BibitemShut {NoStop}%
\bibitem [{\citenamefont {Li}\ \emph {et~al.}(2009)\citenamefont {Li},
  \citenamefont {Zhang}, \citenamefont {Yao},\ and\ \citenamefont
  {Meng}}]{JOUR}%
  \BibitemOpen
  \bibfield  {author} {\bibinfo {author} {\bibfnamefont {J.}~\bibnamefont
  {Li}}, \bibinfo {author} {\bibfnamefont {Y.}~\bibnamefont {Zhang}}, \bibinfo
  {author} {\bibfnamefont {J.}~\bibnamefont {Yao}}, \ and\ \bibinfo {author}
  {\bibfnamefont {J.}~\bibnamefont {Meng}},\ }\href {\doibase
  10.1007/s11433-009-0194-y} {\bibfield  {journal} {\bibinfo  {journal} {Sci.
  China G}\ }\textbf {\bibinfo {volume} {52}},\ \bibinfo {pages} {1586}
  (\bibinfo {year} {2009})}\BibitemShut {NoStop}%
\bibitem [{\citenamefont {Zhao}\ \emph {et~al.}(2020)\citenamefont {Zhao},
  \citenamefont {Nik\ifmmode \check{s}\else \v{s}\fi{}i\ifmmode~\acute{c}\else
  \'{c}\fi{}},\ and\ \citenamefont {Vretenar}}]{PhysRevC.102.054606}%
  \BibitemOpen
  \bibfield  {author} {\bibinfo {author} {\bibfnamefont {J.}~\bibnamefont
  {Zhao}}, \bibinfo {author} {\bibfnamefont {T.}~\bibnamefont {Nik\ifmmode
  \check{s}\else \v{s}\fi{}i\ifmmode~\acute{c}\else \'{c}\fi{}}}, \ and\
  \bibinfo {author} {\bibfnamefont {D.}~\bibnamefont {Vretenar}},\ }\href
  {\doibase 10.1103/PhysRevC.102.054606} {\bibfield  {journal} {\bibinfo
  {journal} {Phys. Rev. C}\ }\textbf {\bibinfo {volume} {102}},\ \bibinfo
  {pages} {054606} (\bibinfo {year} {2020})}\BibitemShut {NoStop}%
\bibitem [{\citenamefont {Negele}\ \emph {et~al.}(1986)\citenamefont {Negele},
  \citenamefont {Serot}, \citenamefont {Vogt},\ and\ \citenamefont
  {Walecka}}]{osti5443763}%
  \BibitemOpen
  \bibfield  {author} {\bibinfo {author} {\bibfnamefont {J.~W.}\ \bibnamefont
  {Negele}}, \bibinfo {author} {\bibfnamefont {B.~D.}\ \bibnamefont {Serot}},
  \bibinfo {author} {\bibfnamefont {E.}~\bibnamefont {Vogt}}, \ and\ \bibinfo
  {author} {\bibfnamefont {J.~D.}\ \bibnamefont {Walecka}},\ }\href
  {https://www.osti.gov/biblio/5443763} {\ \textbf {\bibinfo {volume} {57}}
  (\bibinfo {year} {1986})}\BibitemShut {NoStop}%
\bibitem [{\citenamefont {Williams}\ \emph {et~al.}(1972)\citenamefont
  {Williams}, \citenamefont {Chan},\ and\ \citenamefont
  {Huizenga}}]{WILLIAMS1972225}%
  \BibitemOpen
  \bibfield  {author} {\bibinfo {author} {\bibfnamefont {F.}~\bibnamefont
  {Williams}}, \bibinfo {author} {\bibfnamefont {G.}~\bibnamefont {Chan}}, \
  and\ \bibinfo {author} {\bibfnamefont {J.}~\bibnamefont {Huizenga}},\ }\href
  {\doibase https://doi.org/10.1016/0375-9474(72)90576-3} {\bibfield  {journal}
  {\bibinfo  {journal} {Nucl. Phys. A}\ }\textbf {\bibinfo {volume} {187}},\
  \bibinfo {pages} {225} (\bibinfo {year} {1972})}\BibitemShut {NoStop}%
\bibitem [{\citenamefont {Bethe}(1936{\natexlab{b}})}]{PhysRev.50.332}%
  \BibitemOpen
  \bibfield  {author} {\bibinfo {author} {\bibfnamefont {H.~A.}\ \bibnamefont
  {Bethe}},\ }\href {\doibase 10.1103/PhysRev.50.332} {\bibfield  {journal}
  {\bibinfo  {journal} {Phys. Rev.}\ }\textbf {\bibinfo {volume} {50}},\
  \bibinfo {pages} {332} (\bibinfo {year} {1936}{\natexlab{b}})}\BibitemShut
  {NoStop}%
\bibitem [{\citenamefont {Døssing}\ and\ \citenamefont
  {Jensen}(1974)}]{DOSSING1974493}%
  \BibitemOpen
  \bibfield  {author} {\bibinfo {author} {\bibfnamefont {T.}~\bibnamefont
  {Døssing}}\ and\ \bibinfo {author} {\bibfnamefont {A.}~\bibnamefont
  {Jensen}},\ }\href {\doibase https://doi.org/10.1016/0375-9474(74)90334-0}
  {\bibfield  {journal} {\bibinfo  {journal} {Nucl. Phys. A}\ }\textbf
  {\bibinfo {volume} {222}},\ \bibinfo {pages} {493} (\bibinfo {year}
  {1974})}\BibitemShut {NoStop}%
\bibitem [{\citenamefont {Hilaire}\ and\ \citenamefont
  {Goriely}(2006)}]{HILAIRE200663}%
  \BibitemOpen
  \bibfield  {author} {\bibinfo {author} {\bibfnamefont {S.}~\bibnamefont
  {Hilaire}}\ and\ \bibinfo {author} {\bibfnamefont {S.}~\bibnamefont
  {Goriely}},\ }\href {\doibase
  https://doi.org/10.1016/j.nuclphysa.2006.08.014} {\bibfield  {journal}
  {\bibinfo  {journal} {Nucl. Phys. A}\ }\textbf {\bibinfo {volume} {779}},\
  \bibinfo {pages} {63} (\bibinfo {year} {2006})}\BibitemShut {NoStop}%
\bibitem [{\citenamefont {Capote}\ \emph {et~al.}(2009)\citenamefont {Capote},
  \citenamefont {Herman}, \citenamefont {Obložinský}, \citenamefont {Young},
  \citenamefont {Goriely}, \citenamefont {Belgya}, \citenamefont {Ignatyuk},
  \citenamefont {Koning}, \citenamefont {Hilaire}, \citenamefont {Plujko},
  \citenamefont {Avrigeanu}, \citenamefont {Bersillon}, \citenamefont
  {Chadwick}, \citenamefont {Fukahori}, \citenamefont {Ge}, \citenamefont
  {Han}, \citenamefont {Kailas}, \citenamefont {Kopecky}, \citenamefont
  {Maslov}, \citenamefont {Reffo}, \citenamefont {Sin}, \citenamefont
  {Soukhovitskii},\ and\ \citenamefont {Talou}}]{1}%
  \BibitemOpen
  \bibfield  {author} {\bibinfo {author} {\bibfnamefont {R.}~\bibnamefont
  {Capote}}, \bibinfo {author} {\bibfnamefont {M.}~\bibnamefont {Herman}},
  \bibinfo {author} {\bibfnamefont {P.}~\bibnamefont {Obložinský}}, \bibinfo
  {author} {\bibfnamefont {P.}~\bibnamefont {Young}}, \bibinfo {author}
  {\bibfnamefont {S.}~\bibnamefont {Goriely}}, \bibinfo {author} {\bibfnamefont
  {T.}~\bibnamefont {Belgya}}, \bibinfo {author} {\bibfnamefont
  {A.}~\bibnamefont {Ignatyuk}}, \bibinfo {author} {\bibfnamefont
  {A.}~\bibnamefont {Koning}}, \bibinfo {author} {\bibfnamefont
  {S.}~\bibnamefont {Hilaire}}, \bibinfo {author} {\bibfnamefont
  {V.}~\bibnamefont {Plujko}}, \bibinfo {author} {\bibfnamefont
  {M.}~\bibnamefont {Avrigeanu}}, \bibinfo {author} {\bibfnamefont
  {O.}~\bibnamefont {Bersillon}}, \bibinfo {author} {\bibfnamefont
  {M.}~\bibnamefont {Chadwick}}, \bibinfo {author} {\bibfnamefont
  {T.}~\bibnamefont {Fukahori}}, \bibinfo {author} {\bibfnamefont
  {Z.}~\bibnamefont {Ge}}, \bibinfo {author} {\bibfnamefont {Y.}~\bibnamefont
  {Han}}, \bibinfo {author} {\bibfnamefont {S.}~\bibnamefont {Kailas}},
  \bibinfo {author} {\bibfnamefont {J.}~\bibnamefont {Kopecky}}, \bibinfo
  {author} {\bibfnamefont {V.}~\bibnamefont {Maslov}}, \bibinfo {author}
  {\bibfnamefont {G.}~\bibnamefont {Reffo}}, \bibinfo {author} {\bibfnamefont
  {M.}~\bibnamefont {Sin}}, \bibinfo {author} {\bibfnamefont {E.}~\bibnamefont
  {Soukhovitskii}}, \ and\ \bibinfo {author} {\bibfnamefont {P.}~\bibnamefont
  {Talou}},\ }\href {\doibase https://doi.org/10.1016/j.nds.2009.10.004}
  {\bibfield  {journal} {\bibinfo  {journal} {Nucl. Data Sheets}\ }\textbf
  {\bibinfo {volume} {110}},\ \bibinfo {pages} {3107} (\bibinfo {year}
  {2009})},\ \bibinfo {note} {special Issue on Nuclear Reaction
  Data}\BibitemShut {NoStop}%
\bibitem [{\citenamefont {Koning}\ and\ \citenamefont
  {Rochman}(2012)}]{KONING20122841}%
  \BibitemOpen
  \bibfield  {author} {\bibinfo {author} {\bibfnamefont {A.}~\bibnamefont
  {Koning}}\ and\ \bibinfo {author} {\bibfnamefont {D.}~\bibnamefont
  {Rochman}},\ }\href {\doibase https://doi.org/10.1016/j.nds.2012.11.002}
  {\bibfield  {journal} {\bibinfo  {journal} {Nucl. Data Sheets}\ }\textbf
  {\bibinfo {volume} {113}},\ \bibinfo {pages} {2841} (\bibinfo {year}
  {2012})},\ \bibinfo {note} {special Issue on Nuclear Reaction
  Data}\BibitemShut {NoStop}%
\bibitem [{\citenamefont {Larsen}\ \emph {et~al.}(2013)\citenamefont {Larsen},
  \citenamefont {Ruud}, \citenamefont {B\"urger}, \citenamefont {Goriely},
  \citenamefont {Guttormsen}, \citenamefont {G\"orgen}, \citenamefont {Hagen},
  \citenamefont {Harissopulos}, \citenamefont {Nyhus}, \citenamefont
  {Renstr\o{}m}, \citenamefont {Schiller}, \citenamefont {Siem}, \citenamefont
  {Tveten}, \citenamefont {Voinov},\ and\ \citenamefont
  {Wiedeking}}]{PhysRevC.87.014319}%
  \BibitemOpen
  \bibfield  {author} {\bibinfo {author} {\bibfnamefont {A.~C.}\ \bibnamefont
  {Larsen}}, \bibinfo {author} {\bibfnamefont {I.~E.}\ \bibnamefont {Ruud}},
  \bibinfo {author} {\bibfnamefont {A.}~\bibnamefont {B\"urger}}, \bibinfo
  {author} {\bibfnamefont {S.}~\bibnamefont {Goriely}}, \bibinfo {author}
  {\bibfnamefont {M.}~\bibnamefont {Guttormsen}}, \bibinfo {author}
  {\bibfnamefont {A.}~\bibnamefont {G\"orgen}}, \bibinfo {author}
  {\bibfnamefont {T.~W.}\ \bibnamefont {Hagen}}, \bibinfo {author}
  {\bibfnamefont {S.}~\bibnamefont {Harissopulos}}, \bibinfo {author}
  {\bibfnamefont {H.~T.}\ \bibnamefont {Nyhus}}, \bibinfo {author}
  {\bibfnamefont {T.}~\bibnamefont {Renstr\o{}m}}, \bibinfo {author}
  {\bibfnamefont {A.}~\bibnamefont {Schiller}}, \bibinfo {author}
  {\bibfnamefont {S.}~\bibnamefont {Siem}}, \bibinfo {author} {\bibfnamefont
  {G.~M.}\ \bibnamefont {Tveten}}, \bibinfo {author} {\bibfnamefont
  {A.}~\bibnamefont {Voinov}}, \ and\ \bibinfo {author} {\bibfnamefont
  {M.}~\bibnamefont {Wiedeking}},\ }\href {\doibase 10.1103/PhysRevC.87.014319}
  {\bibfield  {journal} {\bibinfo  {journal} {Phys. Rev. C}\ }\textbf {\bibinfo
  {volume} {87}},\ \bibinfo {pages} {014319} (\bibinfo {year}
  {2013})}\BibitemShut {NoStop}%
\bibitem [{\citenamefont {Guttormsen}\ \emph {et~al.}(2003)\citenamefont
  {Guttormsen}, \citenamefont {Bagheri}, \citenamefont {Chankova},
  \citenamefont {Rekstad}, \citenamefont {Siem}, \citenamefont {Schiller},\
  and\ \citenamefont {Voinov}}]{PhysRevC.68.064306}%
  \BibitemOpen
  \bibfield  {author} {\bibinfo {author} {\bibfnamefont {M.}~\bibnamefont
  {Guttormsen}}, \bibinfo {author} {\bibfnamefont {A.}~\bibnamefont {Bagheri}},
  \bibinfo {author} {\bibfnamefont {R.}~\bibnamefont {Chankova}}, \bibinfo
  {author} {\bibfnamefont {J.}~\bibnamefont {Rekstad}}, \bibinfo {author}
  {\bibfnamefont {S.}~\bibnamefont {Siem}}, \bibinfo {author} {\bibfnamefont
  {A.}~\bibnamefont {Schiller}}, \ and\ \bibinfo {author} {\bibfnamefont
  {A.}~\bibnamefont {Voinov}},\ }\href {\doibase 10.1103/PhysRevC.68.064306}
  {\bibfield  {journal} {\bibinfo  {journal} {Phys. Rev. C}\ }\textbf {\bibinfo
  {volume} {68}},\ \bibinfo {pages} {064306} (\bibinfo {year}
  {2003})}\BibitemShut {NoStop}%
\bibitem [{\citenamefont {Long}\ \emph {et~al.}(2004)\citenamefont {Long},
  \citenamefont {Meng}, \citenamefont {Giai},\ and\ \citenamefont
  {Zhou}}]{1999}%
  \BibitemOpen
  \bibfield  {author} {\bibinfo {author} {\bibfnamefont {W.}~\bibnamefont
  {Long}}, \bibinfo {author} {\bibfnamefont {J.}~\bibnamefont {Meng}}, \bibinfo
  {author} {\bibfnamefont {N.~V.}\ \bibnamefont {Giai}}, \ and\ \bibinfo
  {author} {\bibfnamefont {S.-G.}\ \bibnamefont {Zhou}},\ }\href {\doibase
  10.1103/PhysRevC.69.034319} {\bibfield  {journal} {\bibinfo  {journal} {Phys.
  Rev. C}\ }\textbf {\bibinfo {volume} {69}},\ \bibinfo {pages} {034319}
  (\bibinfo {year} {2004})}\BibitemShut {NoStop}%
\bibitem [{\citenamefont {Lalazissis}\ \emph {et~al.}(2009)\citenamefont
  {Lalazissis}, \citenamefont {Karatzikos}, \citenamefont {Fossion},
  \citenamefont {Arteaga}, \citenamefont {Afanasjev},\ and\ \citenamefont
  {Ring}}]{LALAZISSIS200936}%
  \BibitemOpen
  \bibfield  {author} {\bibinfo {author} {\bibfnamefont {G.}~\bibnamefont
  {Lalazissis}}, \bibinfo {author} {\bibfnamefont {S.}~\bibnamefont
  {Karatzikos}}, \bibinfo {author} {\bibfnamefont {R.}~\bibnamefont {Fossion}},
  \bibinfo {author} {\bibfnamefont {D.~P.}\ \bibnamefont {Arteaga}}, \bibinfo
  {author} {\bibfnamefont {A.}~\bibnamefont {Afanasjev}}, \ and\ \bibinfo
  {author} {\bibfnamefont {P.}~\bibnamefont {Ring}},\ }\href {\doibase
  https://doi.org/10.1016/j.physletb.2008.11.070} {\bibfield  {journal}
  {\bibinfo  {journal} {Phys. Lett. B}\ }\textbf {\bibinfo {volume} {671}},\
  \bibinfo {pages} {36} (\bibinfo {year} {2009})}\BibitemShut {NoStop}%
\bibitem [{\citenamefont {Nikšić}\ \emph {et~al.}(2014)\citenamefont
  {Nikšić}, \citenamefont {Paar}, \citenamefont {Vretenar},\ and\
  \citenamefont {Ring}}]{NIKSIC20141808}%
  \BibitemOpen
  \bibfield  {author} {\bibinfo {author} {\bibfnamefont {T.}~\bibnamefont
  {Nikšić}}, \bibinfo {author} {\bibfnamefont {N.}~\bibnamefont {Paar}},
  \bibinfo {author} {\bibfnamefont {D.}~\bibnamefont {Vretenar}}, \ and\
  \bibinfo {author} {\bibfnamefont {P.}~\bibnamefont {Ring}},\ }\href {\doibase
  https://doi.org/10.1016/j.cpc.2014.02.027} {\bibfield  {journal} {\bibinfo
  {journal} {Computer Physics Communications}\ }\textbf {\bibinfo {volume}
  {185}},\ \bibinfo {pages} {1808} (\bibinfo {year} {2014})}\BibitemShut
  {NoStop}%
\bibitem [{\citenamefont {Lalazissis}\ \emph {et~al.}(2005)\citenamefont
  {Lalazissis}, \citenamefont {Nik\ifmmode \check{s}\else
  \v{s}\fi{}i\ifmmode~\acute{c}\else \'{c}\fi{}}, \citenamefont {Vretenar},\
  and\ \citenamefont {Ring}}]{PhysRevC.71.024312}%
  \BibitemOpen
  \bibfield  {author} {\bibinfo {author} {\bibfnamefont {G.~A.}\ \bibnamefont
  {Lalazissis}}, \bibinfo {author} {\bibfnamefont {T.}~\bibnamefont
  {Nik\ifmmode \check{s}\else \v{s}\fi{}i\ifmmode~\acute{c}\else \'{c}\fi{}}},
  \bibinfo {author} {\bibfnamefont {D.}~\bibnamefont {Vretenar}}, \ and\
  \bibinfo {author} {\bibfnamefont {P.}~\bibnamefont {Ring}},\ }\href {\doibase
  10.1103/PhysRevC.71.024312} {\bibfield  {journal} {\bibinfo  {journal} {Phys.
  Rev. C}\ }\textbf {\bibinfo {volume} {71}},\ \bibinfo {pages} {024312}
  (\bibinfo {year} {2005})}\BibitemShut {NoStop}%
\bibitem [{\citenamefont {Litvinova}\ and\ \citenamefont
  {Afanasjev}(2011)}]{PhysRevC.84.014305}%
  \BibitemOpen
  \bibfield  {author} {\bibinfo {author} {\bibfnamefont {E.~V.}\ \bibnamefont
  {Litvinova}}\ and\ \bibinfo {author} {\bibfnamefont {A.~V.}\ \bibnamefont
  {Afanasjev}},\ }\href {\doibase 10.1103/PhysRevC.84.014305} {\bibfield
  {journal} {\bibinfo  {journal} {Phys. Rev. C}\ }\textbf {\bibinfo {volume}
  {84}},\ \bibinfo {pages} {014305} (\bibinfo {year} {2011})}\BibitemShut
  {NoStop}%
\bibitem [{\citenamefont {Ramirez}\ \emph {et~al.}(2013)\citenamefont
  {Ramirez}, \citenamefont {Voinov}, \citenamefont {Grimes}, \citenamefont
  {Schiller}, \citenamefont {Brune}, \citenamefont {Massey},\ and\
  \citenamefont {Salas-Bacci}}]{PhysRevC.88.064324}%
  \BibitemOpen
  \bibfield  {author} {\bibinfo {author} {\bibfnamefont {A.~P.~D.}\
  \bibnamefont {Ramirez}}, \bibinfo {author} {\bibfnamefont {A.~V.}\
  \bibnamefont {Voinov}}, \bibinfo {author} {\bibfnamefont {S.~M.}\
  \bibnamefont {Grimes}}, \bibinfo {author} {\bibfnamefont {A.}~\bibnamefont
  {Schiller}}, \bibinfo {author} {\bibfnamefont {C.~R.}\ \bibnamefont {Brune}},
  \bibinfo {author} {\bibfnamefont {T.~N.}\ \bibnamefont {Massey}}, \ and\
  \bibinfo {author} {\bibfnamefont {A.}~\bibnamefont {Salas-Bacci}},\ }\href
  {\doibase 10.1103/PhysRevC.88.064324} {\bibfield  {journal} {\bibinfo
  {journal} {Phys. Rev. C}\ }\textbf {\bibinfo {volume} {88}},\ \bibinfo
  {pages} {064324} (\bibinfo {year} {2013})}\BibitemShut {NoStop}%
\bibitem [{\citenamefont {Chankova}\ \emph {et~al.}(2006)\citenamefont
  {Chankova}, \citenamefont {Schiller}, \citenamefont {Agvaanluvsan},
  \citenamefont {Algin}, \citenamefont {Bernstein}, \citenamefont {Guttormsen},
  \citenamefont {Ingebretsen}, \citenamefont {L\"onnroth}, \citenamefont
  {Messelt}, \citenamefont {Mitchell}, \citenamefont {Rekstad}, \citenamefont
  {Siem}, \citenamefont {Larsen}, \citenamefont {Voinov},\ and\ \citenamefont
  {\O{}deg\aa{}rd}}]{PhysRevC.73.034311}%
  \BibitemOpen
  \bibfield  {author} {\bibinfo {author} {\bibfnamefont {R.}~\bibnamefont
  {Chankova}}, \bibinfo {author} {\bibfnamefont {A.}~\bibnamefont {Schiller}},
  \bibinfo {author} {\bibfnamefont {U.}~\bibnamefont {Agvaanluvsan}}, \bibinfo
  {author} {\bibfnamefont {E.}~\bibnamefont {Algin}}, \bibinfo {author}
  {\bibfnamefont {L.~A.}\ \bibnamefont {Bernstein}}, \bibinfo {author}
  {\bibfnamefont {M.}~\bibnamefont {Guttormsen}}, \bibinfo {author}
  {\bibfnamefont {F.}~\bibnamefont {Ingebretsen}}, \bibinfo {author}
  {\bibfnamefont {T.}~\bibnamefont {L\"onnroth}}, \bibinfo {author}
  {\bibfnamefont {S.}~\bibnamefont {Messelt}}, \bibinfo {author} {\bibfnamefont
  {G.~E.}\ \bibnamefont {Mitchell}}, \bibinfo {author} {\bibfnamefont
  {J.}~\bibnamefont {Rekstad}}, \bibinfo {author} {\bibfnamefont
  {S.}~\bibnamefont {Siem}}, \bibinfo {author} {\bibfnamefont {A.~C.}\
  \bibnamefont {Larsen}}, \bibinfo {author} {\bibfnamefont {A.}~\bibnamefont
  {Voinov}}, \ and\ \bibinfo {author} {\bibfnamefont {S.}~\bibnamefont
  {\O{}deg\aa{}rd}},\ }\href {\doibase 10.1103/PhysRevC.73.034311} {\bibfield
  {journal} {\bibinfo  {journal} {Phys. Rev. C}\ }\textbf {\bibinfo {volume}
  {73}},\ \bibinfo {pages} {034311} (\bibinfo {year} {2006})}\BibitemShut
  {NoStop}%
\bibitem [{\citenamefont {Syed}\ \emph {et~al.}(2009)\citenamefont {Syed},
  \citenamefont {Guttormsen}, \citenamefont {Ingebretsen}, \citenamefont
  {Larsen}, \citenamefont {L\"onnroth}, \citenamefont {Rekstad}, \citenamefont
  {Schiller}, \citenamefont {Siem},\ and\ \citenamefont
  {Voinov}}]{PhysRevC.79.024316}%
  \BibitemOpen
  \bibfield  {author} {\bibinfo {author} {\bibfnamefont {N.~U.~H.}\
  \bibnamefont {Syed}}, \bibinfo {author} {\bibfnamefont {M.}~\bibnamefont
  {Guttormsen}}, \bibinfo {author} {\bibfnamefont {F.}~\bibnamefont
  {Ingebretsen}}, \bibinfo {author} {\bibfnamefont {A.~C.}\ \bibnamefont
  {Larsen}}, \bibinfo {author} {\bibfnamefont {T.}~\bibnamefont {L\"onnroth}},
  \bibinfo {author} {\bibfnamefont {J.}~\bibnamefont {Rekstad}}, \bibinfo
  {author} {\bibfnamefont {A.}~\bibnamefont {Schiller}}, \bibinfo {author}
  {\bibfnamefont {S.}~\bibnamefont {Siem}}, \ and\ \bibinfo {author}
  {\bibfnamefont {A.}~\bibnamefont {Voinov}},\ }\href {\doibase
  10.1103/PhysRevC.79.024316} {\bibfield  {journal} {\bibinfo  {journal} {Phys.
  Rev. C}\ }\textbf {\bibinfo {volume} {79}},\ \bibinfo {pages} {024316}
  (\bibinfo {year} {2009})}\BibitemShut {NoStop}%
\bibitem [{\citenamefont {Melby}\ \emph {et~al.}(2001)\citenamefont {Melby},
  \citenamefont {Guttormsen}, \citenamefont {Rekstad}, \citenamefont
  {Schiller}, \citenamefont {Siem},\ and\ \citenamefont
  {Voinov}}]{PhysRevC.63.044309}%
  \BibitemOpen
  \bibfield  {author} {\bibinfo {author} {\bibfnamefont {E.}~\bibnamefont
  {Melby}}, \bibinfo {author} {\bibfnamefont {M.}~\bibnamefont {Guttormsen}},
  \bibinfo {author} {\bibfnamefont {J.}~\bibnamefont {Rekstad}}, \bibinfo
  {author} {\bibfnamefont {A.}~\bibnamefont {Schiller}}, \bibinfo {author}
  {\bibfnamefont {S.}~\bibnamefont {Siem}}, \ and\ \bibinfo {author}
  {\bibfnamefont {A.}~\bibnamefont {Voinov}},\ }\href {\doibase
  10.1103/PhysRevC.63.044309} {\bibfield  {journal} {\bibinfo  {journal} {Phys.
  Rev. C}\ }\textbf {\bibinfo {volume} {63}},\ \bibinfo {pages} {044309}
  (\bibinfo {year} {2001})}\BibitemShut {NoStop}%
\bibitem [{\citenamefont {Rahmatinejad}\ \emph {et~al.}(2020)\citenamefont
  {Rahmatinejad}, \citenamefont {Shneidman}, \citenamefont {Antonenko},
  \citenamefont {Bezbakh}, \citenamefont {Adamian},\ and\ \citenamefont
  {Malov}}]{PhysRevC.101.054315}%
  \BibitemOpen
  \bibfield  {author} {\bibinfo {author} {\bibfnamefont {A.}~\bibnamefont
  {Rahmatinejad}}, \bibinfo {author} {\bibfnamefont {T.~M.}\ \bibnamefont
  {Shneidman}}, \bibinfo {author} {\bibfnamefont {N.~V.}\ \bibnamefont
  {Antonenko}}, \bibinfo {author} {\bibfnamefont {A.~N.}\ \bibnamefont
  {Bezbakh}}, \bibinfo {author} {\bibfnamefont {G.~G.}\ \bibnamefont
  {Adamian}}, \ and\ \bibinfo {author} {\bibfnamefont {L.~A.}\ \bibnamefont
  {Malov}},\ }\href {\doibase 10.1103/PhysRevC.101.054315} {\bibfield
  {journal} {\bibinfo  {journal} {Phys. Rev. C}\ }\textbf {\bibinfo {volume}
  {101}},\ \bibinfo {pages} {054315} (\bibinfo {year} {2020})}\BibitemShut
  {NoStop}%
\bibitem [{\citenamefont {Larsen}\ \emph {et~al.}(2006)\citenamefont {Larsen},
  \citenamefont {Chankova}, \citenamefont {Guttormsen}, \citenamefont
  {Ingebretsen}, \citenamefont {Messelt}, \citenamefont {Rekstad},
  \citenamefont {Siem}, \citenamefont {Syed}, \citenamefont {\O{}deg\aa{}rd},
  \citenamefont {L\"onnroth}, \citenamefont {Schiller},\ and\ \citenamefont
  {Voinov}}]{PhysRevC.73.064301}%
  \BibitemOpen
  \bibfield  {author} {\bibinfo {author} {\bibfnamefont {A.~C.}\ \bibnamefont
  {Larsen}}, \bibinfo {author} {\bibfnamefont {R.}~\bibnamefont {Chankova}},
  \bibinfo {author} {\bibfnamefont {M.}~\bibnamefont {Guttormsen}}, \bibinfo
  {author} {\bibfnamefont {F.}~\bibnamefont {Ingebretsen}}, \bibinfo {author}
  {\bibfnamefont {S.}~\bibnamefont {Messelt}}, \bibinfo {author} {\bibfnamefont
  {J.}~\bibnamefont {Rekstad}}, \bibinfo {author} {\bibfnamefont
  {S.}~\bibnamefont {Siem}}, \bibinfo {author} {\bibfnamefont {N.~U.~H.}\
  \bibnamefont {Syed}}, \bibinfo {author} {\bibfnamefont {S.~W.}\ \bibnamefont
  {\O{}deg\aa{}rd}}, \bibinfo {author} {\bibfnamefont {T.}~\bibnamefont
  {L\"onnroth}}, \bibinfo {author} {\bibfnamefont {A.}~\bibnamefont
  {Schiller}}, \ and\ \bibinfo {author} {\bibfnamefont {A.}~\bibnamefont
  {Voinov}},\ }\href {\doibase 10.1103/PhysRevC.73.064301} {\bibfield
  {journal} {\bibinfo  {journal} {Phys. Rev. C}\ }\textbf {\bibinfo {volume}
  {73}},\ \bibinfo {pages} {064301} (\bibinfo {year} {2006})}\BibitemShut
  {NoStop}%
\bibitem [{\citenamefont {Alhassan}\ \emph {et~al.}(2022)\citenamefont
  {Alhassan}, \citenamefont {Rochman}, \citenamefont {Vasiliev}, \citenamefont
  {Hursin}, \citenamefont {Koning},\ and\ \citenamefont
  {Ferroukhi}}]{Alhassan2022}%
  \BibitemOpen
  \bibfield  {author} {\bibinfo {author} {\bibfnamefont {E.}~\bibnamefont
  {Alhassan}}, \bibinfo {author} {\bibfnamefont {D.}~\bibnamefont {Rochman}},
  \bibinfo {author} {\bibfnamefont {A.}~\bibnamefont {Vasiliev}}, \bibinfo
  {author} {\bibfnamefont {M.}~\bibnamefont {Hursin}}, \bibinfo {author}
  {\bibfnamefont {A.~J.}\ \bibnamefont {Koning}}, \ and\ \bibinfo {author}
  {\bibfnamefont {H.}~\bibnamefont {Ferroukhi}},\ }\href {\doibase
  https://doi.org/10.1007/s41365-022-01034-w} {\bibfield  {journal} {\bibinfo
  {journal} {Nucl. Sci. Tech}\ }\textbf {\bibinfo {volume} {33}},\ \bibinfo
  {pages} {50} (\bibinfo {year} {2022})}\BibitemShut {NoStop}%
\bibitem [{\citenamefont {Toft}\ \emph {et~al.}(2010)\citenamefont {Toft},
  \citenamefont {Larsen}, \citenamefont {Agvaanluvsan}, \citenamefont
  {B\"urger}, \citenamefont {Guttormsen}, \citenamefont {Mitchell},
  \citenamefont {Nyhus}, \citenamefont {Schiller}, \citenamefont {Siem},
  \citenamefont {Syed},\ and\ \citenamefont {Voinov}}]{PhysRevC.81.064311}%
  \BibitemOpen
  \bibfield  {author} {\bibinfo {author} {\bibfnamefont {H.~K.}\ \bibnamefont
  {Toft}}, \bibinfo {author} {\bibfnamefont {A.~C.}\ \bibnamefont {Larsen}},
  \bibinfo {author} {\bibfnamefont {U.}~\bibnamefont {Agvaanluvsan}}, \bibinfo
  {author} {\bibfnamefont {A.}~\bibnamefont {B\"urger}}, \bibinfo {author}
  {\bibfnamefont {M.}~\bibnamefont {Guttormsen}}, \bibinfo {author}
  {\bibfnamefont {G.~E.}\ \bibnamefont {Mitchell}}, \bibinfo {author}
  {\bibfnamefont {H.~T.}\ \bibnamefont {Nyhus}}, \bibinfo {author}
  {\bibfnamefont {A.}~\bibnamefont {Schiller}}, \bibinfo {author}
  {\bibfnamefont {S.}~\bibnamefont {Siem}}, \bibinfo {author} {\bibfnamefont
  {N.~U.~H.}\ \bibnamefont {Syed}}, \ and\ \bibinfo {author} {\bibfnamefont
  {A.}~\bibnamefont {Voinov}},\ }\href {\doibase 10.1103/PhysRevC.81.064311}
  {\bibfield  {journal} {\bibinfo  {journal} {Phys. Rev. C}\ }\textbf {\bibinfo
  {volume} {81}},\ \bibinfo {pages} {064311} (\bibinfo {year}
  {2010})}\BibitemShut {NoStop}%
\bibitem [{\citenamefont {Agvaanluvsan}\ \emph {et~al.}(2004)\citenamefont
  {Agvaanluvsan}, \citenamefont {Schiller}, \citenamefont {Becker},
  \citenamefont {Bernstein}, \citenamefont {Garrett}, \citenamefont
  {Guttormsen}, \citenamefont {Mitchell}, \citenamefont {Rekstad},
  \citenamefont {Siem}, \citenamefont {Voinov},\ and\ \citenamefont
  {Younes}}]{PhysRevC.70.054611}%
  \BibitemOpen
  \bibfield  {author} {\bibinfo {author} {\bibfnamefont {U.}~\bibnamefont
  {Agvaanluvsan}}, \bibinfo {author} {\bibfnamefont {A.}~\bibnamefont
  {Schiller}}, \bibinfo {author} {\bibfnamefont {J.~A.}\ \bibnamefont
  {Becker}}, \bibinfo {author} {\bibfnamefont {L.~A.}\ \bibnamefont
  {Bernstein}}, \bibinfo {author} {\bibfnamefont {P.~E.}\ \bibnamefont
  {Garrett}}, \bibinfo {author} {\bibfnamefont {M.}~\bibnamefont {Guttormsen}},
  \bibinfo {author} {\bibfnamefont {G.~E.}\ \bibnamefont {Mitchell}}, \bibinfo
  {author} {\bibfnamefont {J.}~\bibnamefont {Rekstad}}, \bibinfo {author}
  {\bibfnamefont {S.}~\bibnamefont {Siem}}, \bibinfo {author} {\bibfnamefont
  {A.}~\bibnamefont {Voinov}}, \ and\ \bibinfo {author} {\bibfnamefont
  {W.}~\bibnamefont {Younes}},\ }\href {\doibase 10.1103/PhysRevC.70.054611}
  {\bibfield  {journal} {\bibinfo  {journal} {Phys. Rev. C}\ }\textbf {\bibinfo
  {volume} {70}},\ \bibinfo {pages} {054611} (\bibinfo {year}
  {2004})}\BibitemShut {NoStop}%
\bibitem [{\citenamefont {Ramirez}\ \emph {et~al.}(2015)\citenamefont
  {Ramirez}, \citenamefont {Voinov}, \citenamefont {Grimes}, \citenamefont
  {Byun}, \citenamefont {Brune}, \citenamefont {Massey}, \citenamefont
  {Akhtar}, \citenamefont {Dhakal},\ and\ \citenamefont
  {Parker}}]{PhysRevC.92.014303}%
  \BibitemOpen
  \bibfield  {author} {\bibinfo {author} {\bibfnamefont {A.~P.~D.}\
  \bibnamefont {Ramirez}}, \bibinfo {author} {\bibfnamefont {A.~V.}\
  \bibnamefont {Voinov}}, \bibinfo {author} {\bibfnamefont {S.~M.}\
  \bibnamefont {Grimes}}, \bibinfo {author} {\bibfnamefont {Y.}~\bibnamefont
  {Byun}}, \bibinfo {author} {\bibfnamefont {C.~R.}\ \bibnamefont {Brune}},
  \bibinfo {author} {\bibfnamefont {T.~N.}\ \bibnamefont {Massey}}, \bibinfo
  {author} {\bibfnamefont {S.}~\bibnamefont {Akhtar}}, \bibinfo {author}
  {\bibfnamefont {S.}~\bibnamefont {Dhakal}}, \ and\ \bibinfo {author}
  {\bibfnamefont {C.~E.}\ \bibnamefont {Parker}},\ }\href {\doibase
  10.1103/PhysRevC.92.014303} {\bibfield  {journal} {\bibinfo  {journal} {Phys.
  Rev. C}\ }\textbf {\bibinfo {volume} {92}},\ \bibinfo {pages} {014303}
  (\bibinfo {year} {2015})}\BibitemShut {NoStop}%
\bibitem [{\citenamefont {Oginni}\ \emph {et~al.}(2009)\citenamefont {Oginni},
  \citenamefont {Grimes}, \citenamefont {Voinov}, \citenamefont {Adekola},
  \citenamefont {Brune}, \citenamefont {Carter}, \citenamefont {Heinen},
  \citenamefont {Jacobs}, \citenamefont {Massey}, \citenamefont {O'Donnell},\
  and\ \citenamefont {Schiller}}]{PhysRevC.80.034305}%
  \BibitemOpen
  \bibfield  {author} {\bibinfo {author} {\bibfnamefont {B.~M.}\ \bibnamefont
  {Oginni}}, \bibinfo {author} {\bibfnamefont {S.~M.}\ \bibnamefont {Grimes}},
  \bibinfo {author} {\bibfnamefont {A.~V.}\ \bibnamefont {Voinov}}, \bibinfo
  {author} {\bibfnamefont {A.~S.}\ \bibnamefont {Adekola}}, \bibinfo {author}
  {\bibfnamefont {C.~R.}\ \bibnamefont {Brune}}, \bibinfo {author}
  {\bibfnamefont {D.~E.}\ \bibnamefont {Carter}}, \bibinfo {author}
  {\bibfnamefont {Z.}~\bibnamefont {Heinen}}, \bibinfo {author} {\bibfnamefont
  {D.}~\bibnamefont {Jacobs}}, \bibinfo {author} {\bibfnamefont {T.~N.}\
  \bibnamefont {Massey}}, \bibinfo {author} {\bibfnamefont {J.~E.}\
  \bibnamefont {O'Donnell}}, \ and\ \bibinfo {author} {\bibfnamefont
  {A.}~\bibnamefont {Schiller}},\ }\href {\doibase 10.1103/PhysRevC.80.034305}
  {\bibfield  {journal} {\bibinfo  {journal} {Phys. Rev. C}\ }\textbf {\bibinfo
  {volume} {80}},\ \bibinfo {pages} {034305} (\bibinfo {year}
  {2009})}\BibitemShut {NoStop}%
\end{thebibliography}%

\end{document}